\documentclass[aps,twocolumn]{revtex4}
\pdfoutput=1

\usepackage{graphicx}
\usepackage{times}
\usepackage{amsmath}
\usepackage{amssymb}

\begin{document}

\title{Adhesion of membranes {\em via} receptor-ligand complexes: \\ Domain formation, binding cooperativity, and active processes}
\author{Thomas R. Weikl}
\author{Mesfin Asfaw}\altaffiliation[Present address: ]{Asia-Pacific Center for Theoretical Physics, Pohang 790-784, Korea}
\author{Heinrich Krobath}
\author{ Bartosz R\'{o}\.{z}ycki}
  \altaffiliation[Present address: ]{Laboratory of Chemical Physics, National Institute of Diabetes and Digestive and Kidney Diseases, National Institutes of Health, Bethesda, MD 20892-0520, USA}
\author{Reinhard Lipowsky}
\affiliation{Max Planck Institute of Colloids and Interfaces, Department of Theory and Bio-Systems, 14424 Potsdam, Germany}

\begin{abstract}
Cell membranes interact {\em via} anchored receptor and ligand molecules. Central questions on cell adhesion concern the binding affinity of these membrane-anchored molecules, the mechanisms leading to the receptor-ligand domains observed during adhesion, and the role of cytoskeletal and other active processes. In this review, these questions are addressed from a theoretical perspective. We focus on models in which the membranes are described as elastic sheets, and the receptors and ligands as anchored molecules. 
In these models, the thermal membrane roughness on the nanometer scale leads to a cooperative binding of anchored receptor and ligand molecules, since the receptor-ligand binding smoothens out the membranes and facilitates the formation of additional bonds. Patterns of receptor domains observed in Monte Carlo simulations point towards a joint role of spontaneous and active processes in cell adhesion. The interactions mediated by the receptors and ligand molecules can be characterized by effective membrane adhesion potentials that depend on the concentrations and binding energies of the molecules.
\end{abstract}

\maketitle

\section{Introduction}
\label{section_introduction}

The adhesion of cells is mediated by the specific binding of receptor and ligand molecules anchored in the cell membranes. Cell adhesion processes are essential for the distinction of self and foreign in immune responses, the formation of tissues, or the signal transduction across the synaptic cleft of neurons \cite{Alberts02}. These processes have therefore been studied intensively with a variety of experimental methods \cite{Alon95,Grakoui99,Delanoe04,Arnold04,Mossman05}. In addition, experiments on lipid vesicles with membrane-anchored receptor and ligand molecules aim to mimic the specific membrane binding processes leading to cell adhesion \cite{Albersdoerfer97,Maier01,Smith08}.

In many adhesion processes, the anchored receptor and ligand molecules can still diffuse, at least to some extent, within the contact area of the adhering membranes \cite{Grakoui99,Delanoe04,Mossman05}. As a consequence, the receptor-ligand complexes may form different spatial patterns such as clusters or extended domains in the contact area \cite{Grakoui99,Mossman05,Monks98,Davis04}. These pattern formation processes can be understood in the framework of discrete models in which the membranes are divided into small patches, and the receptors and ligands are described as single molecules that are either present or absent in the patches \cite{Lipowsky96,Weikl00,Weikl04,Asfaw06,Weikl06}. These discrete models are lattice models on elastic surfaces, and have two advantages: (i) They automatically incorporate the mutual exclusion of receptor or ligand  molecules anchored within the same membrane; and (ii) they lead to effective membrane adhesion potentials that provide an intuitive understanding of the observed behavior in terms  of nucleation and growth processes.

Cell adhesion involves many different length scales, from nanometers to tens of micrometers. The largest length scales of micrometers correspond to the diameter of the cell and the diameter of the contact zone in which the cell is bound to another cell or to a supported membrane. The separation of the two membranes in the cell contact zone is orders of magnitude smaller. The membrane separation is comparable to the length of the receptor-ligand complexes, which have a typical extension between 15 and 40 nanometers \cite{Dustin00}. Finally, the smallest relevant length scale is the binding range of a receptor and a ligand molecule, i.e.~the difference between the smallest and the largest local membrane separation at which the molecules can bind. The binding range reflects (i) the range of the lock-and-key interaction, (ii) the flexibility of the two binding partners, and (iii) the flexibility of the membrane anchoring. For the rather rigid protein receptors and ligands that typically mediate cell adhesion, the interaction range is around 1 nanometer. In contrast, the interaction range of surface-anchored flexible tether molecules with specific binding sites is significantly larger \cite{Jeppesen01,Morreira03,Moore06,Martin06}.

The wide range of length scales has important consequences for modeling cell adhesion. In general, the elasticity of the cell membrane is affected by the bending rigidity $\kappa$ of the membrane \cite{Helfrich73}, the membrane tension $\sigma$, and the cytoskeleton that is coupled to the membrane \cite{Gov03,Fournier04,Lin04,Auth07}. The tension dominates over the bending elasticity on length scales larger than the crossover length $\sqrt{\kappa/\sigma}$ \cite{Lipowsky95}, which is of the order of several hundred nanometers for cell membranes \cite{Krobath07}, while the bending elasticity dominates on length scales smaller than the crossover length. The elastic contribution of the cytoskeleton is relevant on length scales larger than the average distance of the cytoskeletal membrane anchors, which is around 100 nanometers \cite{Alberts02}.  The overall shape of the cell membrane on micrometer scales therefore is governed by the cytoskeletal elasticity and membrane tension. In the cell adhesion zone, however, the relevant shape deformations and fluctuations of the membranes occur on length scales up to the average distance of the receptor-ligand bonds, since the bonds locally constrain the membrane separation. The average distance of the bonds roughly varies between 50 and 100 nanometers for typical bond concentrations in cell adhesion zones  \cite{Grakoui99}. The relevant membrane shape deformations and fluctuations in the cell contact zone are therefore dominated by the bending rigidity of the membranes.

The adhesion of cells is mediated by a multitude of different receptor and ligand molecules. Some of these molecules can be strongly coupled to the cytoskeleton. In focal adhesions of cells, for example, clusters of integrin molecules are tightly coupled to the cytoskeleton {\em via} supramolecular assemblies that impose constraints on the lateral separation of the integrins \cite{Arnold04,Selhuber08}. Through focal adhesions, cells exert and sense forces \cite{Geiger01,Discher05,Bershadsky06,Girard07,Schwarz07,De08}. Other receptor and ligand molecules are not \cite{Delanoe04} or only weakly \cite{DeMond08} coupled to the cytoskeleton. These molecules are mobile and diffuse within the membranes. The diffusion process can be observed with single-molecule experiments \cite{Schuetz00,Sako00}.

The adhesion of membranes {\em via} mobile receptor and ligand molecules has been studied theoretically with a variety of models. These models can be grouped into two classes. In both classes, the membranes are described as thin elastic sheets. In the first class of models, the description is continuous in space, and the distribution of the membrane-anchored receptor and ligand molecules on the membranes are given by continuous concentration profiles \cite{Bell78,Bell84,Komura00,Bruinsma00,Chen03,Coombs04,Shenoy05,Wu06}. Dynamic, time-dependent properties of such models have been studied by numerical solution of reaction-diffusion equations \cite{Qi01,Raychaudhuri03,Burroughs02,Shenoy05}. In the second, more recent class of models, the membranes are discretized, and the receptors and ligands are described as single molecules \cite{Lipowsky96,Weikl00,Weikl01,Weikl02b,Asfaw06,Weikl06,Smith05,Krobath07,Tsourkas07,Tsourkas08,Reister08}. The dynamic properties can be numerically studied with Monte Carlo simulations \cite{Weikl02a,Weikl04,Krobath07,Tsourkas07,Tsourkas08}, and central aspects of the equilibrium behavior can be directly inferred from the partition function \cite{Weikl00,Weikl01,Weikl06,Asfaw06}.

\section{Effective adhesion potential}
\label{section_effective_adhesion_potential}

\begin{figure}[t]
\begin{center}
\resizebox{\columnwidth}{!}{\includegraphics{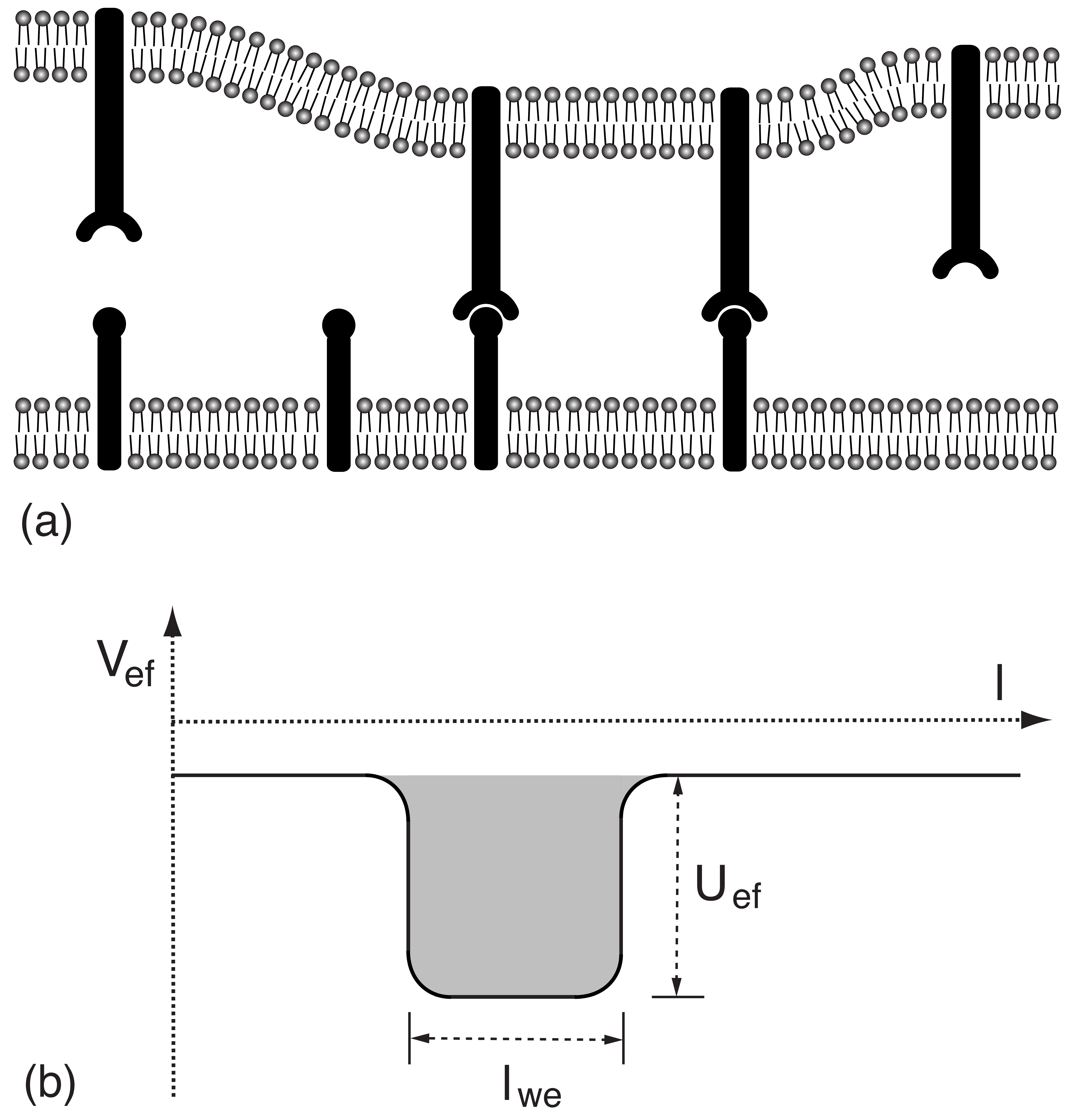}}
\caption{(a) A membrane segment with receptor molecules (top) interacting with ligands embedded in an apposing membrane (bottom). A receptor can bind a ligand molecule if the local separation of the membranes is close to the length of the receptor-ligand complex. -- (b) The attractive interactions between the receptor and ligand molecules lead to an effective single-well adhesion potential $V_\text{ef}$ of the membranes. The depth $U_\text{ef}$ of the potential well depends on the concentrations and binding affinity of receptors and ligands, see eq.~(\ref{Uef}). The width $l_\text{we}$ of the binding well is equal to the binding range of the receptor-ligand interaction.
}
\label{figure_model_one}
\end{center}
\end{figure}

In discrete models, the two apposing membranes in the contact zone of cells or vesicles are divided into small patches \cite{Lipowsky96,Weikl00,Weikl01,Weikl02b,Asfaw06,Weikl06,Smith05,Krobath07,Tsourkas07,Tsourkas08,Reister08}. These patches can contain a single receptor or ligand molecule. Mobile receptor and ligand molecules diffuse by `hopping' from patch to patch, and the thermal fluctuations of the membranes are reflected in variations of the local separation of apposing membrane patches. A receptor can bind to a ligand molecule if the ligand is located in the membrane patch apposing the receptor, and if the local separation of the membranes is close to the length of the receptor-ligand complex. In these models, the linear size $a$ of the membrane patches is typically chosen around 5 nm to capture the whole spectrum of bending deformations of the lipid membranes \cite{Goetz99}.  

Cells can interact {\em via} a multitude of different receptors and ligands. However, it is instructive to start with the relatively simple situation in which the adhesion is mediated by a single type of receptor-ligand bonds as in fig.~\ref{figure_model_one}(a). Such a situation occurs if a cell adheres to a supported membane with a single type of ligands, or if a vesicle with membrane-anchored receptors adheres to a membrane with complementary ligands. The effective membrane adhesion potential mediated by the receptor and ligands can be calculated by integrating over all possible positions of the receptors and ligands in the partition function of the models  \cite{Weikl06,Weikl01}. In the case of a single type of receptors and ligands, the effective adhesion potential of the membranes is a single-well potential with the same range $l_\text{we}$ as the receptor-ligand interaction, but an effective binding energy $U_\text{ef}$ that depends on the concentrations and binding energy $U$ of receptors and ligands, see fig.~\ref{figure_model_one}(b). For typical concentrations of receptors and ligands in cell membranes, which are more than two orders of magnitude smaller than the maximum concentration $1/a^2\simeq 4\cdot 10^4/ \mu\text{m}^2$ in our discretized membranes with patch size $a\simeq 5$ nm, the effective potential depth is \cite{KrobathPreprint}
\begin{equation}
U_{\rm ef} \approx k_B T \, [R] [L ]\, a^2 e^{U/k_BT} 
\label{Uef}
\end{equation}
where $[R]$ and $[L]$ are the area concentrations of unbound receptors and ligands. The quantity
\begin{equation}
K_\text{pl} \equiv a^2 e^{U/k_BT}
\label{Kpl}
\end{equation}
in eq.~(\ref{Uef}) can be interpreted as the binding equilibrium constant of the receptors and ligands in the case of two planar and parallel membranes with a separation equal to the length of the receptor-ligand bonds. The equilibrium constant characterizes the binding affinity of the molecules and can, in principle, be measured with the surface force apparatus in which the apposing membranes are supported on rigid substrates \cite{Israelachvili92,Bayas07}. In the case of flexible membranes, the binding affinity of the receptors and ligands is more difficult to capture, see next section.

\begin{figure}[t]
\begin{center}
\resizebox{\columnwidth}{!}{\includegraphics{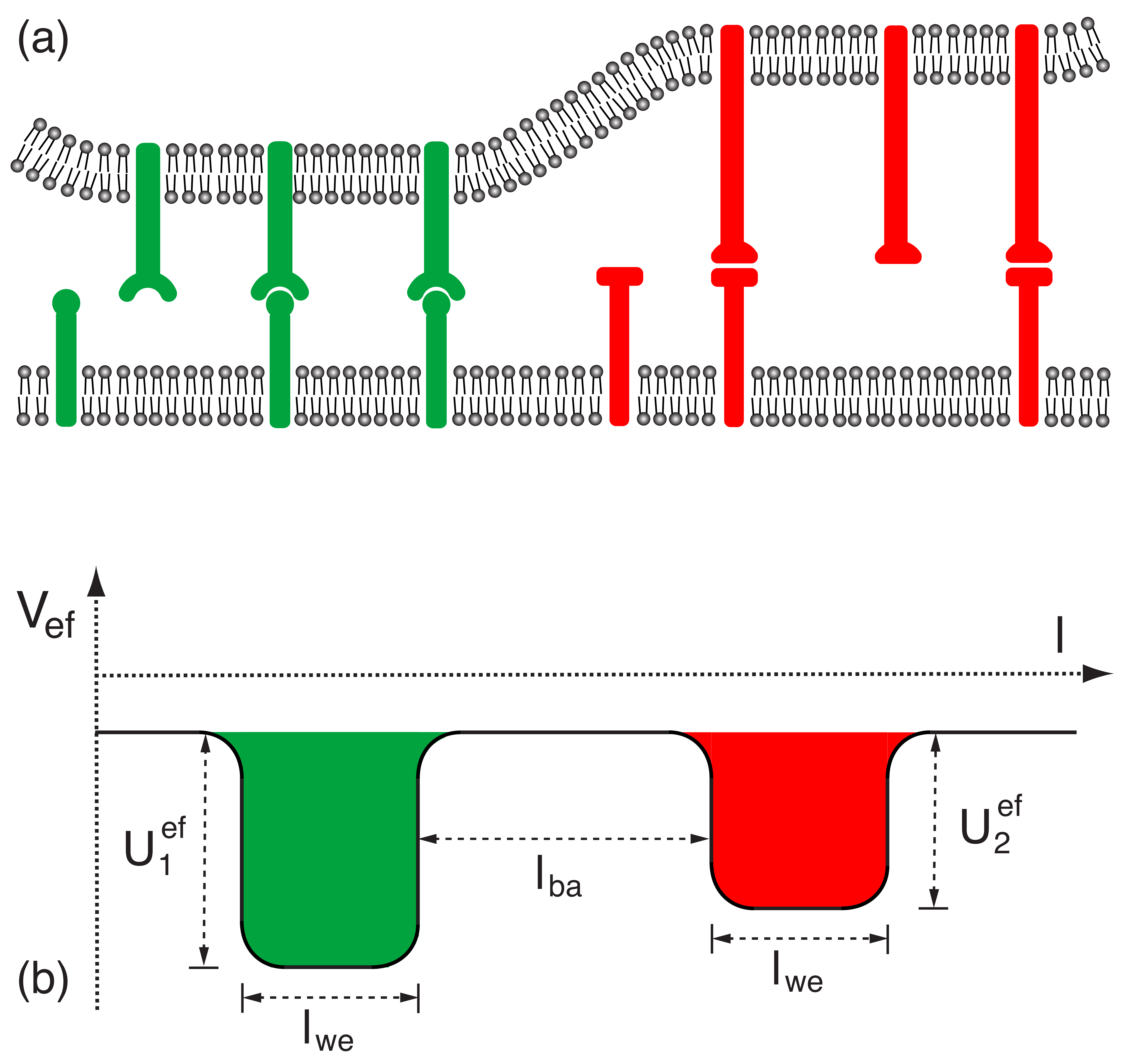}}
\caption{(a) Two membranes  interacting {\it via} long (red) and short (green) receptor-ligand complexes. -- (b) The attractive interactions between the two types of receptors and ligands lead to an effective double-well adhesion potential $V_\text{ef}$ of the membranes. The potential well 1 at small membrane separations $l$ reflects the interactions of the short receptor-ligand complexes, and the potential well 2 at larger membrane separations the interactions of the long receptor-ligand complexes. The depths $U_1^\text{ef}$ and $U_2^\text{ef}$ of the two potential wells depend on the concentrations and binding energies of the two types of receptors and ligands, see eqs.~(\ref{U1ef}) and (\ref{U2ef}). }
\label{figure_model_two}
\end{center}
\end{figure}

The interaction of cells is often mediated by several types of receptor-ligand complexes that differ in their length. For two types of receptors and ligands as in fig.~\ref{figure_model_two}, the effective adhesion potential of the membranes is a double-well potential \cite{Asfaw06}. The depths of the two wells 
\begin{eqnarray}
U_1^\text{ef} \approx k_BT \, [R_1][L_1] \, a^2 e^{U_1/k_BT} 
\label{U1ef}\\
U_2^\text{ef} \approx k_BT \, [R_2][L_2] \, a^2 e^{U_2/k_BT} 
\label{U2ef}
\end{eqnarray}
depend on the concentrations and binding energies $U_1$ and $U_2$ of the different types of receptors and ligands \cite{KrobathPreprint}. In analogy to eq.~(\ref{Kpl}), the quantities $K_\text{pl,1}\equiv a^2 e^{U_1/k_BT}$ and $K_\text{pl,2}\equiv a^2 e^{U_2/k_BT}$ can be interpreted as binding equilibrium constants in the case of planar membranes with a separation equal to the lengths $l_1$ or $l_2$ of the receptor-ligand complexes. 

Repulsive membrane-anchored molecules such as anchored polymers or glycoproteins can lead to additional barriers in the effective adhesion potential \cite{Weikl01,Weikl02b,Weikl06}. The effective adhesion potentials simplify the characterization of the equilibrium properties of the membranes, and lead to an intuitive understanding of these properties, see next sections.

\section{Binding cooperativity}
\label{section_binding_cooperativity}

A receptor molecule can only bind an apposing ligand if the local membrane separation is comparable to the length of the receptor-ligand complex. A central quantity therefore is the fraction $P_b$ of the apposing membranes with a separation within the binding range of the receptor-ligand interaction. The concentration of bound receptor-ligand complexes 
\begin{equation}
[RL] \approx P_b\, K_\text{pl}\,  [R] [L] 
\label{RL}
\end{equation}
is proportional to $P_b$ as well as to the concentrations $[R]$ and $[L]$ of unbound receptors and ligands \cite{KrobathPreprint}. 

Thermal shape fluctuations of the membranes on nanometer scales in general lead to values of $P_b$ smaller than 1. For cell membranes, these nanometer scale fluctuations are not, or only weakly, suppressed by the cell cytoskeleton, in contrast to large-scale shape fluctuations. For simplicity, we assume now that the adhesion of the membranes is mediated by a single type of receptor and ligand molecules as in 
fig.~\ref{figure_model_one}(a). The precise value of $P_b$ then depends  on the well depth $U_\text{ef}$ of the effective adhesion potential shown in fig.~\ref{figure_model_one}(b), and on the bending rigidities of the membranes. For typical lengths and concentrations of receptors and ligands in cell adhesion zones, the fraction $P_b$ of the membranes within binding range of the receptors and ligands turns out to be much smaller than 1, and scaling analysis and Monte Carlo simulations lead to the relation \cite{KrobathPreprint}
\begin{equation}
P_b \approx  c \, \kappa\,  l_\text{we}^2 \, U_\text{ef} / (k_B T)^2 
\label{Pb_approx}
\end{equation}
with prefactor $c=13 \pm 1$. With eqs.~(\ref{Uef}) and (\ref{Kpl}), we obtain 
\begin{equation}
P_b \approx c \, (\kappa/k_B T) l_\text{we}^2 K_\text{pl }[R][L] 
\label{Pb_approx_II}
\end{equation}
which shows that the membrane fraction $P_b$ within the binding range of the receptors and ligands is proportional to $[R]$ and $[L]$. Inserting eq.~(\ref{Pb_approx_II}) into eq.~(\ref{RL}) leads to 
\begin{equation}
[RL] \approx c (\kappa/k_B T) l_\text{we}^2 K_\text{pl}^2 [R]^2 [L]^2 
\label{central_equation}
\end{equation}
The concentration $[RL]$ of receptor-ligand complexes in the adhesion zone thus depends quadratically on the concentrations $[R]$ and $[L]$ of unbound receptors and ligands, which indicates cooperative binding. The binding cooperativity results from a `smoothening' of the thermally rough membranes and, thus, an increase of $P_b$ with increasing concentrations $[R]$ and $[L]$ of receptors and ligands, which facilitates the formation of additional receptor-ligand complexes. The relations (\ref{Pb_approx_II}) and (\ref{central_equation}) are good approximations up to $P_b\lesssim 0.2$, and can be extended to larger values of $P_b$ \cite{KrobathPreprint}.

For {\em soluble} receptor and ligand molecules, in contrast, the volume concentration 
of the bound receptor-ligand complexes 
\begin{equation}
[RL]_\text{3D} = K_\text{3D} [R]_\text{3D} [L]_\text{3D}
\label{equilibrium_constant}
\end{equation}
is proportional to the volume concentrations $[R]_\text{3D}$ and $[L]_\text{3D}$ of unbound receptors and unbound ligands in the solution. The binding affinity of the molecules then can be characterized by the equilibrium constant $K_\text{3D}$, which depends on the binding free energy of the complex \cite{Schuck97,Rich00,McDonnell01}. In analogy to eq.~(\ref{equilibrium_constant}), the binding affinity of membrane-anchored receptors and ligands is often quantified by 
\begin{equation}
K_\text{2D} \equiv \frac{[RL]}{[R][L]}
\label{K2D}
\end{equation}
where $[RL]$,  $[R]$, and $[L]$ are the area concentrations of bound receptor-ligand complexes, unbound receptors, and unbound ligands \cite{Orsello01,Dustin01,Williams01}. 

However, it follows from our relation (\ref{central_equation}) that $K_\text{2D}$ is not constant, but depends on the concentrations of the receptors and ligands. From the eqs.~(\ref{Kpl}) and (\ref{RL}), we obtain the general relation
\begin{equation}
K_\text{2D} = P_b K_\text{pl} 
\label{K2DKpl}
\end{equation}
As mentioned in the previous section, $K_\text{pl}$ is the well-defined two-dimensional equilibrium constant of the receptors and ligands in the case of planar membranes with $P_b=1$, e.g.~two supported membranes in the surface force apparatus with a separation equal to the length of the receptor-ligand complex \cite{Israelachvili92,Bayas07}. 

The relation (\ref{K2DKpl}) also helps to understand why different experimental methods for measuring $K_\text{2D}$ have led to values that differ by several orders of magnitude \cite{Dustin01}. In fluorescence recovery experiments, $K_\text{2D}$ is measured in the equilibrated contact zone of a cell adhering to a supported membrane with fluorescently labeled ligands \cite{Dustin96,Dustin97,Zhu07,Tolentino08}. In micropipette experiments, in contrast, $K_\text{2D}$ is measured for initial contacts between two cells \cite{Chesla98,Williams01,Huang04}, for which $P_b$ can be several orders of magnitude smaller than in equilibrium  \cite{KrobathPreprint}.

\section{Adhesion of vesicles}
\label{section_adhesion_of_vesicles}

\begin{figure}[t]
\begin{center}
\resizebox{0.95\columnwidth}{!}{\includegraphics{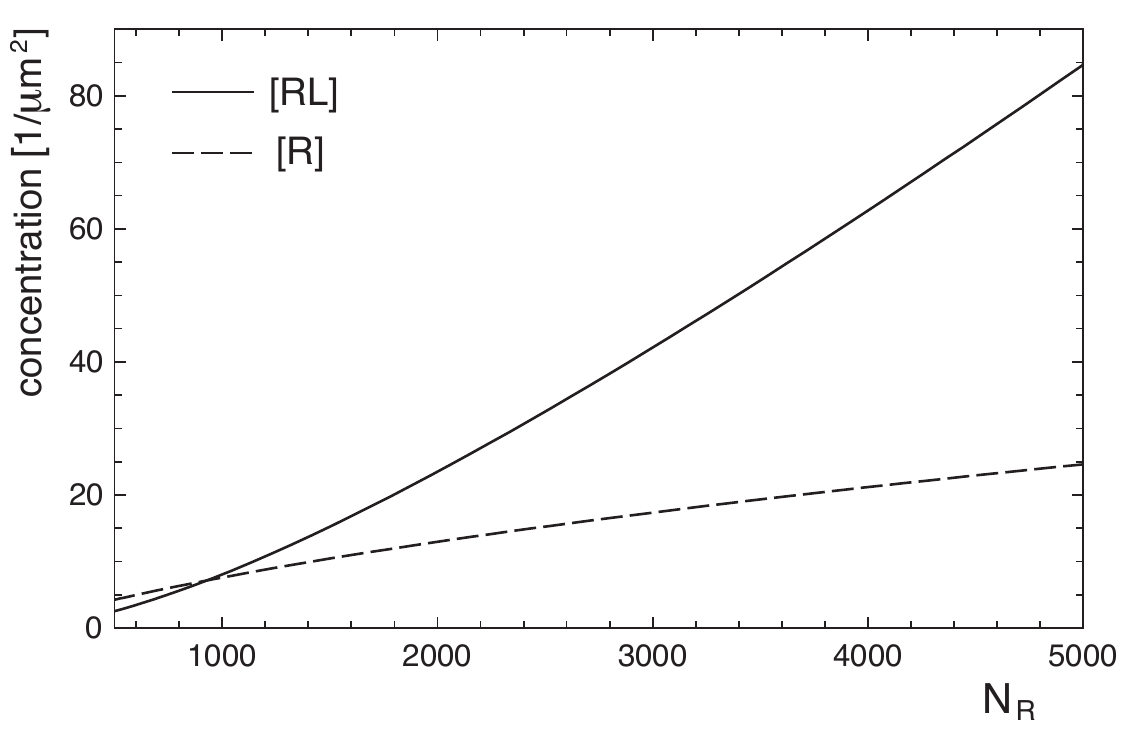}}
\caption{Concentrations $[RL]$ and $[R]$ of bound and unbound receptors as a function of the receptor number $N_R$ of a vesicle adhering to a supported membrane, see eqs.~(\ref{Rvesicle}) and (\ref{RLvesicle}). In this numerical example, we have chosen the bending rigidity $\kappa = 25 k_B T$, the binding range $l_\text{we}=1$ nm, the binding equilibrium constant $K_\text{pl} = 1\, \mu \text{m}^2$ for planar membranes, and a ligand concentration $[L] = 20/ \mu \text{m}^2$ for the supported membrane, which results in the value $0.14/\mu \text{m}^2$ for the parameter $b$ in eqs.~(\ref{Rvesicle}) and (\ref{RLvesicle}). The total area of the vesicle is $100\,  \mu \text{m}^2$, and the area of the contact zone is $30\,\mu\text{m}^2$. The fraction $P_b$ of the vesicle membrane in the contact zone with a separation within binding range of the receptors and ligands varies with $[R]$, see eq.~\ref{Pb_approx_II}, and attains the maximum value $P_b = 0.17$ for $N_R=5000$ in this example.
}
\label{figure_vesicle}
\end{center}
\end{figure}

Important aspects of cell adhesion can be mimicked by lipid vesicles with anchored receptor molecules \cite{Albersdoerfer97,Kloboucek99,Maier01,Smith06,Lorz07,Purrucker07,Smith08}. We focus here on a vesicle adhering to a supported membrane with complementary ligands. In the strong adhesion limit, the shape of the vesicle can be approximated by a spherical cap \cite{Lipowsky05,Seifert90}. The volume of the cap depends on the osmotic pressure balance between the outside and the interior of the vesicle.
If this volume is nearly constant,  the contact area $A_c$ is nearly independent of the adhesion free energy \cite{Seifert90,Tordeux02,Gruhn05,Smith05}. 

Since the total number $N_R$ of receptors in the vesicle membrane is fixed, we have
\begin{equation}
N_R = [R] A + [RL] A_c
\end{equation}
where $A$ is the total area of the vesicle, and $[RL]$ is the concentration of receptor-ligand complexes in the contact area. For typical small concentrations of receptors and ligands, the concentration $[R]$ of unbound receptors within the contact area and within the non-adhering membrane section of the vesicle are approximately equal in equilibrium since the excluded volume of the receptor-ligand complexes in the contact area is negligible. With eq.~(\ref{central_equation}), we then obtain
\begin{equation}
[R] = \frac{\sqrt{A^2 + 4 b A_c N_R}-A}{2 b A_c}
\label{Rvesicle}
\end{equation}
and
\begin{equation}
[RL] = \frac{\left(\sqrt{A^2 + 4 b A_c N_R}-A\right)^2}{4 b A_c^2}
\label{RLvesicle}
\end{equation}
with $b = c (\kappa/k_B T) l_\text{we}^2 K_\text{pl}^2 [L]^2$. Here, $[L]$ is the concentration of the unbound ligands in the supported membrane, which is nearly independent of the binding  state of the vesicle if the membrane is large. Because of the binding cooperativity of the receptors and ligands, the concentrations $[R]$ and $[RL]$ are not linear in $N_R$, see fig.~\ref{figure_vesicle} for a numerical example. 

\section{Domains of long and short receptor-ligand complexes}

\subsection{Critical concentrations for domain formation}
\label{subsection_critical_concentrations}

\begin{figure*}[t]
\begin{center}
\resizebox{2\columnwidth}{!}{\includegraphics{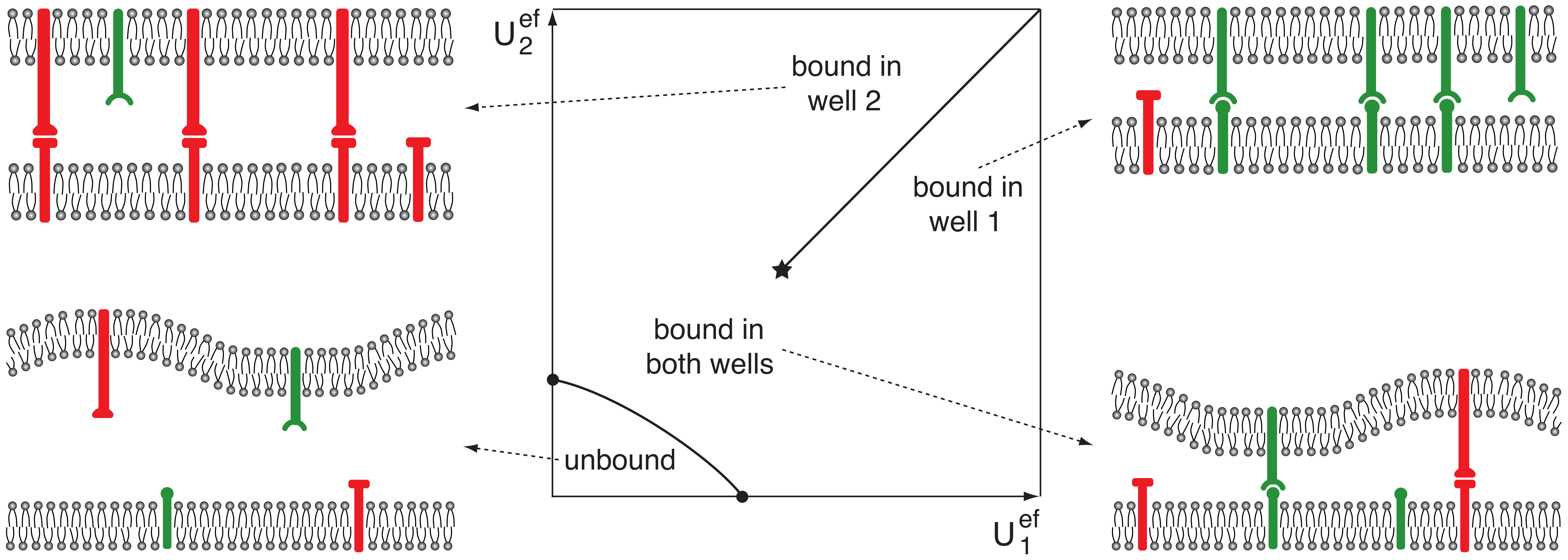}}
\caption{Phase diagram of membranes adhering via long and short receptor/ligand complexes. 
The membranes are unbound for small well depths $U_1^\text{ef}$ and $U_2^\text{ef}$ of the effective interaction potential shown in Fig.~\ref{figure_model_two}(b), i.e.~for small concentrations or binding energies of receptors and ligands, see
eqs.~(\ref{U1ef}) and (\ref{U2ef}). At large values of $U_1^\text{ef}$ and $U_2^\text{ef}$, the membranes are either bound in well 1 or well 2, i.e.~they are either bound by the short or by the long receptor/ligand complexes. At intermediate well depths $U_1^\text{ef}$ and $U_2^\text{ef}$, the membranes are bound in both potential wells. The critical point for the lateral phase separation (star) follows from eq.~(\ref{Uc}). For typical dimensions of cell receptors and ligands, the critical well depth $U_c^\text{ef}$ for lateral phase separation is significantly larger than the critical depths of unbinding \cite{Asfaw06}. In the absence of other repulsive interactions as assumed here, the membranes unbind due to steric repulsion.
 }
\label{figure_phase_diagram}
\end{center} 
\end{figure*}

Cells often interact {\em via} receptor-ligand complexes that differ significantly in size. For example, two important complexes in T-cell adhesion are the complexes of the T-cell receptor (TCR) with a length of 15 nm and integrin complexes with a length of 40 nm \cite{Dustin00}. The length mismatch induces a membrane-mediated repulsion between the different complexes because the membranes have to be curved to compensate the mismatch, which costs bending energy. 

The equilibrium behavior of two membranes adhering {\em via} long and short receptor-ligand complexes is determined by the effective double-well adhesion potential shown in fig.~\ref{figure_model_two}(b), and by the bending rigidities of the membranes. The depths of the two wells reflect the concentrations and binding affinity of the two different types of receptors and ligands, see eqs.~(\ref{U1ef}) and (\ref{U2ef}) in section \ref{section_effective_adhesion_potential}. If the two wells are relatively shallow, membrane segments can easily cross the barrier between the wells, driven by thermal fluctuations. If the two wells are deep, the crossing from one well to the other well is suppressed by the potential barrier of width $l_\text{ba}$ between the wells. The potential barrier induces a line tension between a membrane segment that is bound in one of the wells and an adjacent membrane segment bound in the other well. Beyond a critical depth of the potential wells, the line tension leads to the formation of large membrane domains that are bound in well one or well two, see fig.~\ref{figure_phase_diagram}. Within each domain, the adhesion of the membranes is predominantly mediated by one of the two types of receptor-ligand complexes.

Scaling arguments indicate that domains bound in either well 1 or well 2 are formed if the depths $U_\text{1}^\text{ef}$ and $U_\text{2}^\text{ef}$ of the two potential wells exceed the critical potential depth \cite{Asfaw06}
\begin{equation}
U_c^\text{ef} = \frac{c (k_BT)^2}{\kappa l_\text{we} l_\text{ba}} \label{Uc}
\end{equation}
Numerical results from Monte Carlo simulations confirm eq.~(\ref{Uc}) and lead to the value $c = 0.225 \pm 0.02$ for the dimensionless prefactor \cite{Asfaw06}. The critical potential depth thus depends on the effective rigidity $\kappa=\kappa_1\kappa_2/(\kappa_1+\kappa_2)$ of two membranes with bending  rigidities $\kappa_1$ and $\kappa_2$ and the width $l_\text{we}$ and separation $l_\text{ba}$ of the two potential wells, see fig.~\ref{figure_model_two}(b). The separation $l_\text{ba}$ of the wells is close to the length mismatch of the different types of receptor-ligand complexes, which is 25 nm for T cells. A reasonable estimate for the interaction range $l_\text{we}$ of the protein receptors and ligands that mediate cell adhesion is 1 nm, see section \ref{section_introduction}. With an effective bending rigidity $\kappa=\kappa_1\kappa_2/(\kappa_1+\kappa_2)$ of, e.g., 25 $k_B T$, we obtain the estimate $U_c^\text{ef} \simeq 360\,k_B T/\mu\text{m}^2$ for the critical potential depth of domain formation during T-cell adhesion. For planar-membrane equilibrium constants $K_\text{pl,1}$ and $K_\text{pl,2}$ around $1\,\mu\text{m}^2$, for example, the effective potential depths (\ref{U1ef}) and (\ref{U2ef}) exceed this critical potential depth if the concentrations of unbound receptors and ligands are larger than $20/\mu\text{m}^2$.

\subsection{Domain patterns during immune cell adhesion}
\label{section_domain_patterns}

\begin{figure}[t]
\begin{center}
\resizebox{0.6\columnwidth}{!}{\includegraphics{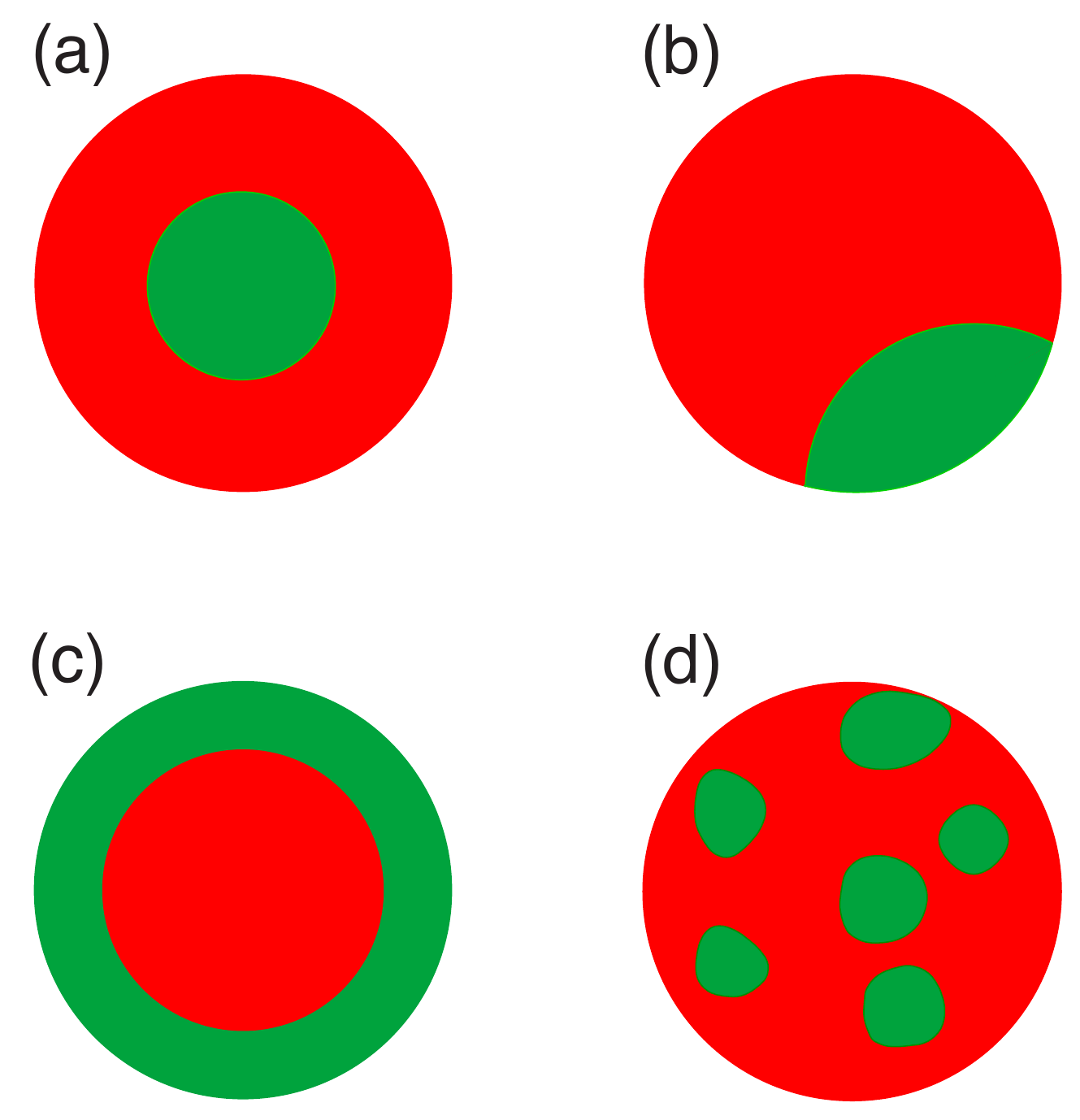}}
\caption{Domain patterns in the T-cell contact zone: (a) Final pattern of helper T cells with a central TCR domain (green) surrounded by an integrin domain (red) \cite{Monks98,Grakoui99}. The pattern results from cytoskeletal transport of TCRs towards the contact zone center \cite{Mossman05,Weikl04}. -- (b) Simulated final pattern in the absence of TCR transport \cite{Weikl04}. The length of the boundary line between the TCR and the integrin domain is minimal in this pattern. -- (c) and (d) The two types of intermediate patterns observed in the first minutes of adhesion \cite{Davis04}. In simulations, both patterns result from the nucleation of TCR clusters in the first seconds of adhesion and the subsequent diffusion of unbound TCR and MHC-peptide ligands in the contact zone \cite{Weikl04}. The closed TCR ring in pattern (c) forms from fast-growing TCR clusters in the periphery of the contact zone at sufficiently large TCR-MHC-peptide concentrations. The pattern (d) forms at smaller TCR-MHC-peptide concentrations.}
\label{figure_patterns}
\end{center}
\end{figure}

The domains of long and short receptor-ligand complexes formed during the adhesion of T cells and other immune cells such as natural killer cells evolve in characteristic patterns. For T cells, the domains either contain complexes of TCR and its ligand  MHC-peptide, or integrin complexes. The final domain pattern in the T-cell contact zone is formed within 15 to 30 minutes and consists of a central TCR domain surrounded by a ring-shaped integrin domain \cite{Monks98,Grakoui99}, see fig.~\ref{figure_patterns}(a). Interestingly, the intermediate patterns formed within the first minutes of T-cell adhesion are quite different \cite{Grakoui99,Davis04}. They are either inverse to the final pattern, with a central integrin domain surrounded by a ring-shaped TCR domain, see fig.~\ref{figure_patterns}(c), or exhibit several nearly circular TCR domains in the contact zone, see fig.~\ref{figure_patterns}(d).

To understand these patterns, several groups have modeled and simulated the adhesion of T cells and other immune cells \cite{Qi01,Weikl02a,Burroughs02,Lee03,Raychaudhuri03,Weikl04,Coombs04,Figge06,Tsourkas07,Tsourkas08}. One open question concerned the role of the T-cell cytoskeleton, which polarizes during adhesion, with a focal point in the center of the contact zone \cite{Alberts02,Dustin98}. Some groups have found that the final T-cell pattern with a central TCR domain can emerge independently of cytoskeletal processes \cite{Qi01,Lee03}. In contrast, Monte Carlo simulations of discrete models indicate that the central TCR cluster is only formed if TCR molecules are actively transported by the cytoskeleton towards the center of the contact zone \cite{Weikl04}. The active transport has been simulated by a biased diffusion of TCRs towards the contact zone center, which implies a weak coupling of TCRs to the cytoskeleton. In the absence of active TCR transport, the Monte Carlo simulations lead to the final, equilibrium pattern shown in fig.~\ref{figure_patterns}(b), which minimizes the energy of the boundary line between the TCR and the integrin domain \cite{Weikl04}. In agreement with these simulations, recent T-cell adhesion experiments on patterned substrates reveal cytoskeletal forces that drive the TCRs towards the center of the contact zone \cite{Mossman05,DeMond08}. The experiments  indicate a weak frictional coupling of the TCRs to the cytoskeletal flow \cite{DeMond08}. 

The intermediate patterns formed in the Monte Carlo simulations closely resemble the intermediate immune-cell patterns shown in figs.~\ref{figure_patterns}(c) and (d). In the first seconds of adhesion, the Monte Carlo patterns exhibit small TCR clusters \cite{Weikl04}. In the following seconds, the diffusion of free TCR and MHC-peptide molecules into the contact zone lead to faster growth of TCR clusters close to the periphery of the contact zone \cite{footnote}. For sufficiently large TCR-MHC-peptide concentrations, the peripheral TCR clusters grow into the ring-shaped domain of fig.~\ref{figure_patterns}(c). At smaller TCR-MHC-peptide concentration, the initial clusters evolve into the pattern of fig.~\ref{figure_patterns}(d). In agreement with experimental observations \cite{Davis04}, only these two types of intermediate patterns are formed in the simulations. The simulated patterns emerge spontaneously from the nucleation of TCR clusters and the diffusion of unbound TCR and MHC-peptide into the contact zone.

\subsection{Implications for T-cell activation}
\begin{figure}[b]
\begin{center}
\resizebox{0.75\columnwidth}{!}{\includegraphics{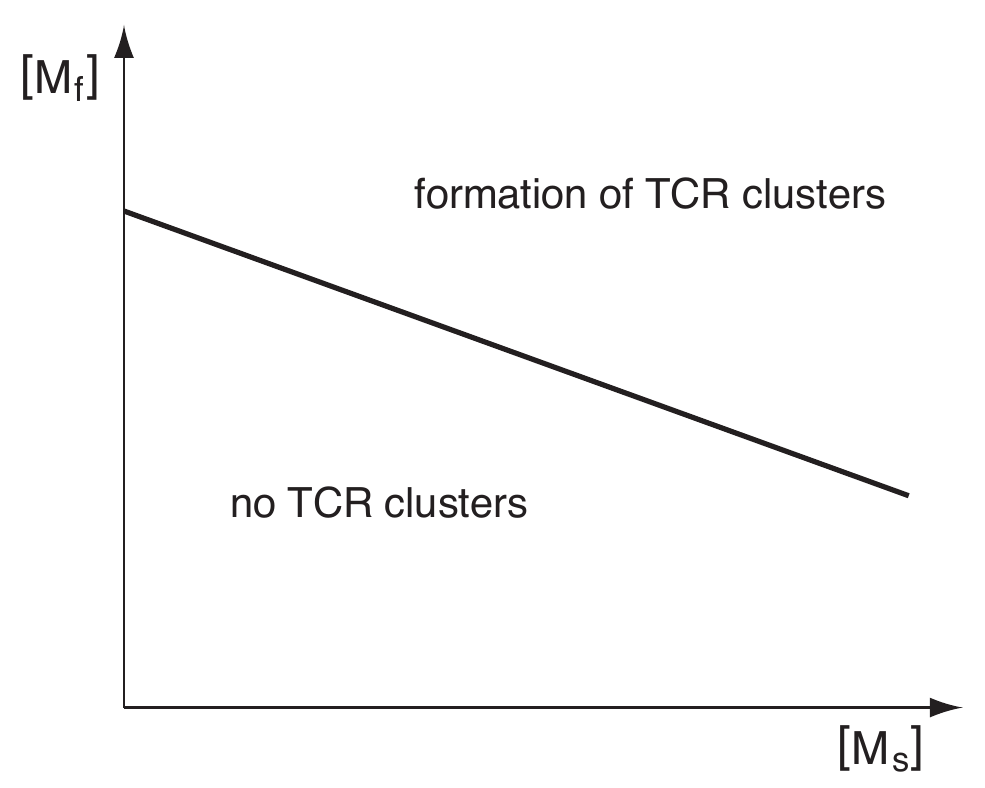}}
\caption{Schematic diagram for the joint role of foreign and self MHC-peptides in TCR cluster formation and T-cell activation. Here, $[M_f]$ is the concentration of foreign MHC-peptides, and $[M_s]$ is the concentration of self MHC-peptides. The solid line represents the threshold for TCR cluster formation given by eq.~(\ref{threshold}). The slope of this line is the negative ratio $K_\text{TMs}/K_\text{TMf}$ of the binding equilibrium constants for the interaction of TCR with self MHC-peptide and with foreign MHC-peptide. For simplicity, we have assumed here a single, dominant type of self MHC-peptides.
}
\label{figure_TCRclusters}
\end{center}
\end{figure}

T cells mediate immune responses by adhering to cells that display foreign peptides on their surfaces \cite{Alberts02,Janeway07}. The peptides are presented by MHC proteins on the cell surfaces, and are recognized by the T-cell receptors (TCR). T-cell activation requires the binding of TCRs to the MHC-peptide complexes. But how precisely these binding events trigger T-cell activation still is a current focus of immunology (for reviews, see refs.~\cite{Choudhuri07,Davis06,Krogsgaard07}). Recent experiments indicate that the first T-cell activation signals coincide with the formation of TCR microclusters within the first seconds of T-cell adhesion \cite{Campi05,Yokosuka05,Bunnell02,Varma06,Yokosuka08,Yokosuka09}. 

In the discrete model introduced in section \ref{section_effective_adhesion_potential} and fig.~\ref{figure_model_two}, TCR clusters in the T-cell contact zone can only form if two conditions are met. First, the effective potential depth $U_1^\text{ef}$ for the short TCR-MHC-peptide complexes and the depth $U_2^\text{ef}$ for the long integrin complexes have to exceed the critical depth (\ref{Uc}). Second, the effective potential $U_1^\text{ef}$ for the TCR complexes has to be larger than the effective depth $U_2^\text{ef}$ in the situation where no TCRs are bound. To understand the second condition, one has to realize that the concentrations of unbound TCRs and unbound integrins depend on the area fractions of the TCR and integrin clusters and domains in the contact zone. If no TCRs are bound, i.e.~if the whole contact zone is occupied by an integrin domain, the concentration of unbound TCRs  is maximal.  Hence, also the effective depth $U_1^\text{ef}$ for the TCRs is maximal in this situation, see eq.~(\ref{U1ef}). TCR clusters now form if $U_1^\text{ef}$ is larger than $U_2^\text{ef}$, which leads to a decrease in the concentration of unbound TCRs and, thus, to a decrease in $U_1^\text{ef}$. The area fraction of the TCR clusters grows until the equilibrium situation with $U_1^\text{ef}= U_2^
\text{ef}$ is reached \cite{Asfaw06}.

T-cell activation requires a threshold concentration of foreign MHC-peptide complexes. Interestingly, the threshold concentration of foreign MHC-peptide depends on the concentration of self MHC-peptide complexes, i.e.~of complexes between MHC and self peptides derived from proteins of the host cell \cite{Irvine02,Purbhoo04}. The foreign MHC-peptide complexes, in contrast, are complexes of MHC with peptides derived from viral or bacterial proteins. Self MHC-peptides typically bind weakly to TCR, since strong binding can result in autoimmune reactions. However, the number of self MHC-peptide complexes typically greatly exceed the number of foreign MHC-peptide on cell surfaces. Both self and foreign MHC-peptide complexes contribute to the effective potential depth $U_1^\text{ef}$ of the TCR-MHC-peptide interaction. For simplicity, we assume here a single, dominant type of self MHC-peptides with concentrations $[M_s]$. The effective potential depth then is
\begin{equation}
U_1^\text{ef} = k_B T \left([T][M_f] K_\text{TMf} + [T][M_s] K_\text{TMs} + \ldots \right)  \label{U1efT}
\end{equation}
where $[T]$ and $[M_f]$ are the concentrations of unbound TCR and foreign MHC-peptide, and $K_\text{TMf}$ and $K_\text{TMs}$ are the binding equilibrium constants of foreign and self TCR-MHC-peptide complexes in the case of planar membranes, see eq.~(\ref{Kpl}). The dots in eq.~(\ref{U1efT}) indicate possible contributions from other receptor-ligand complexes with the same length as the TCR-MHC-peptide complex, e.g.~from the CD2-CD58 complex \cite{Dustin00}. In addition, repulsive glycoproteins with a length larger than TCR-MHC-peptide complex  can affect $U_1^\text{ef}$ \cite{Weikl04,Weikl02a}. Similarly, the depth $U_2^\text{ef}$ of the second well depends on the concentrations and binding equilibrium constants of integrins and its ligands.

Let us now suppose that the numbers of TCRs, co-receptors such as CD2, integrins, and glycoproteins are approximately equal for different T cells and apposing cells, while the numbers of foreign and self MHC-peptides vary. The second condition $U_1^\text{ef}> U_2^\text{ef}$ for TCR cluster formation then leads to 
\begin{equation}
[M_f] K_\text{TMf} + [M_s] K_\text{TMs} > c_t \label{threshold}
\end{equation}
where $c_t$ is a dimensionless threshold that depends on the TCR, co-receptor, integrin, and glycoprotein concentrations, etc.  The threshold concentration of foreign MHC-peptide complexes for TCR cluster formation thus depends on the concentration of self MHC-peptide complexes, see fig.~\ref{figure_TCRclusters}. If the formation of TCR microclusters coincides with early activation signals as suggested in Refs.~\cite{Campi05,Yokosuka05}, the inequality (\ref{threshold}) also helps to understand the joint role of foreign and self MHC-peptides in T-cell activation.

\section{Adhesion {\em via} crosslinker molecules or adsorbed particles}
\label{section_adsorbed_particles}

\begin{figure}[b]
\begin{center}
\resizebox{0.7\columnwidth}{!}{\includegraphics{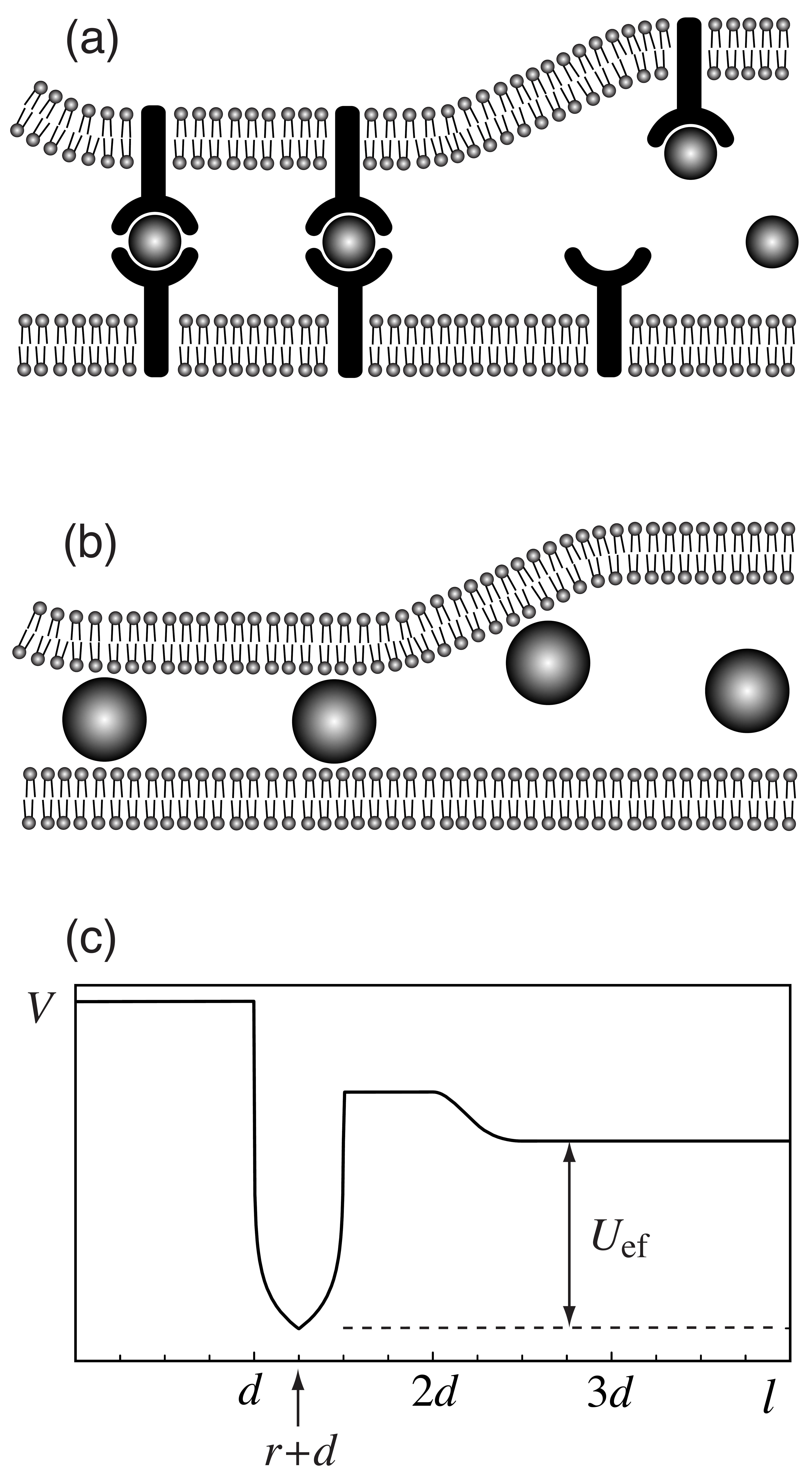}}
\caption{
(a) Two membranes with receptors binding to solute molecules. At small membrane separations, the molecules can bind to two apposing receptors and, thus, crosslink the membranes. -- (b) Two membranes in contact with a solution of adhesive molecules or particles. A particle can bind the two membranes together for membrane separations slightly larger than the particle diameter. At larger separations, the particles can only bind to one of the membranes. -- (c) Effective adhesion potential $V$ of the membranes in subfigure (b) as a function of the membrane separation $l$ for small concentrations of the particles \cite{Rozycki08b}. The effective potential has a minimum at the separation $l=d+r$ where $d$ is the particle diameter, and $r$ is the range of the adhesive interaction between the particles and the membranes. At this separation, the particles are bound to both membranes. The effective potential is constant for large separations at which the particles can only bind to one of the membranes. The potential barrier at intermediate separations $d+ 2r < l < 2d$ results from the fact that a particle bound to one of the membranes locally `blocks' the binding of other particles to the apposing membrane.
}
\label{figure_linkers}
\end{center}
\end{figure}
\begin{figure*}[t]
\begin{center}
\resizebox{2.1\columnwidth}{!}{\includegraphics{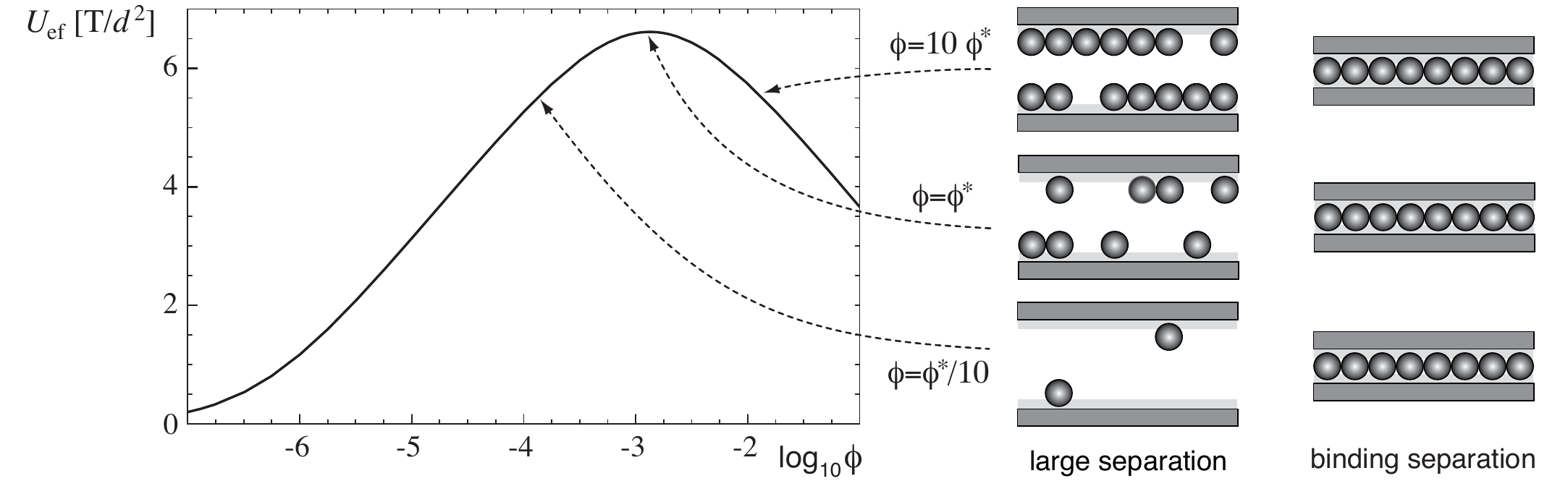}}
\caption{Effective adhesion energy $U_{\rm ef}$, given in eq.~(\ref{Uef_linkers}), as a function of the volume fraction $\phi$ of the adhesive particles for the binding energy $U= 8k_BT$ and $q = 0.25$. The effective adhesion energy is maximal at the optimal volume fraction $\phi^{\star} \approx e^{-U/k_BT}/ q \simeq 1.34 \cdot 10^{-3}$. At the optimal volume fraction, the particle coverage of two planar parallel membranes is close to 50\% for large separations, and almost 100\% for small separations at which the particles can bind to both surfaces \cite{Rozycki08b}. }
\label{figure_linkers_II}
\end{center}
\end{figure*}

The binding of receptor molecules on apposing membranes or surfaces is sometimes mediated by linker or connector molecules, see fig.~\ref{figure_linkers}(a). Biotinylated lipids in apposing membranes, for example, can be crosslinked by the connector molecule streptavidin \cite{Albersdoerfer97,Leckband94}. The effective binding affinity of the membranes then depends both on the area concentrations of the membrane receptors and  the volume concentration of the linker molecules. A similar situation arises if adhesive molecules or particles  directly bind to lipid bilayers \cite{Hu04,Baksh04,Winter06}. The adhesive particles can crosslink two apposing membranes if the membrane separation is close to the particle diameter, see fig.~\ref{figure_linkers}(b). At large membrane separations, the particles can only bind to one of the membranes. 

The effective, particle-mediated adhesion potential of the membranes can be determined by integrating over all possible positions of the adhesive particles or linker molecules in the partition function of the considered model. Conceptually, this is similar to the calculation of the effective adhesion potential for membranes interacting {\em via} anchored receptors and ligands, which requires an integration over all positions of the receptor and ligand molecules in the membranes, see section \ref{section_effective_adhesion_potential}. For simplicity, we consider here the adhesive particles of fig.~\ref{figure_linkers}(b), which interact directly with the lipid bilayers. The explicit integration over the particle positions requires spatial discretizations. In a lattice model, the space between the apposing membranes is discretized into a cubic lattice with a lattice spacing equal to the particle diameter $d$ \cite{Rozycki08b}. In an alternative semi-continuous model, only the two spatial directions parallel to the membranes are discretized, while the third spatial direction perpendicular to the membranes is continuous \cite{Rozycki08b}. In both models, the effective, particle-mediated adhesion potential at large membrane separations has the form  
\begin{equation}
V_\infty \approx - 2\frac{k_B T}{d^2} \ln\left[1+q \phi e^{U/k_B T}\right] 
\label{Vinfty}
\end{equation}
for small volume fractions $\phi$ and large binding energies $U$ of the particles. At small separations close to particle diameter, the adhesion potential exhibits a minimum
\begin{equation}
V_\text{min} \approx - \frac{k_B T}{d^2} \ln\left[1+q \phi e^{2 U/k_B T}\right]
\label{Vmin}
\end{equation}
The model-dependent factor $q$ in eqs.~(\ref{Vinfty}) and (\ref{Vmin}) has the value 1 in the lattice gas model and the value $r/d$ in the semi-continous model with interaction range $r$ of the adhesive particles. In the semi-continuous model, the potential minimum is located at the membrane separation $l=d+r$, see fig.~\ref{figure_linkers}(c). 

Independent of these two models, the eqs.~(\ref{Vinfty}) and (\ref{Vmin}) can also be understood as Langmuir adsorption free energies per binding site. Eq.~(\ref{Vmin}) can be interpreted as the Langmuir adsorption free energy for small membrane separations at which a particle binds both membranes with total binding energy $2U$, and eq.~(\ref{Vinfty}) as the Langmuir adsorption free energy for large separations.  The Langmuir adsorption free energies  result from a simple two-state model in which a particle is either absent (Boltzmann weight $1$) or present (Boltzmann weights $q\, \phi \, e^{2U/k_BT}$ and $q\, \phi \, e^{U/k_BT}$, respectively) at a given binding site.

The effective, particle-mediated adhesion energy of the membranes can be defined as
\begin{equation}
U_\text{ef} \equiv V_\infty - V_\text{min} \approx \frac{k_B T}{d^2} \ln \frac{1+q \phi e^{2 U/k_B T}}{\left(1+q \phi e^{U/k_B T}\right)^2}
\label{Uef_linkers}
\end{equation}
Interestingly, the effective adhesion energy is maximal at the volume fraction $\phi^{\star} \simeq e^{-U/k_BT} / q$, and considerably smaller  at smaller or larger volume fractions, see fig.~\ref{figure_linkers_II}.  At this optimal volume fraction, the particle coverage $c_{\infty} = -(d^2 /2) (\partial V_{\infty} / \partial U) \approx \phi /( \phi + \phi^{\star})$ of the unbound membranes is 50\%. In contrast, the particle coverage $c_{\rm min} = - (d^2 /2) (\partial V_{\rm min} / \partial U) \approx \phi /( \phi + \phi^{\star} e^{-U/k_BT})$  of the bound membranes is close to 100\% at $\phi = \phi^{\star}$. Bringing the surfaces from large separations within binding separations thus does not `require' desorption or adsorption of particles at the optimal volume fraction. The existence of an optimal particle volume fraction has important implications that are accessible to experiments, such as `re-entrant transitions' in which surfaces or colloidal objects first bind with increasing concentration of adhesive particles, and unbind again when the concentration is further increased  beyond the optimal concentration.

\section{Active switching of adhesion receptors}
\label{section_active_switching}

Some adhesion receptors can be switched between different conformations. A biological example of switchable, membrane-anchored adhesion receptors are integrins. In one of their conformations, the integrin molecules are extended and can bind to apposing ligands \cite{Takagi02,Kim03,Dustin04}. In another conformation, the molecules are bent and, thus, deactivated. The transitions between these conformations are triggered by signaling cascades in the cells, which typically require energy input, e.g.~{\em via} ATP.  Because of this energy input, the switching process is an active, non-equilibrium process. In biomimetic applications with designed molecules, active conformational transitions may also be triggered by light \cite{Moeller98,Ichimura00}. Other active processes of biomembranes include the forces exerted by embedded ion pumps \cite{Prost96,Manneville99,Ramaswamy00,Gov04,Lin06,ElAlaouiFaris09} or by the cell cytoskeleton \cite{Gov05,Zhang08b}, see also section \ref{section_domain_patterns}.  Active conformational transitions of membrane proteins have also been suggested to couple to the local thickness \cite{Sabra98} or curvature \cite{Chen04} of the membranes.

\begin{figure}[t]
\begin{center}
\resizebox{\columnwidth}{!}{\includegraphics{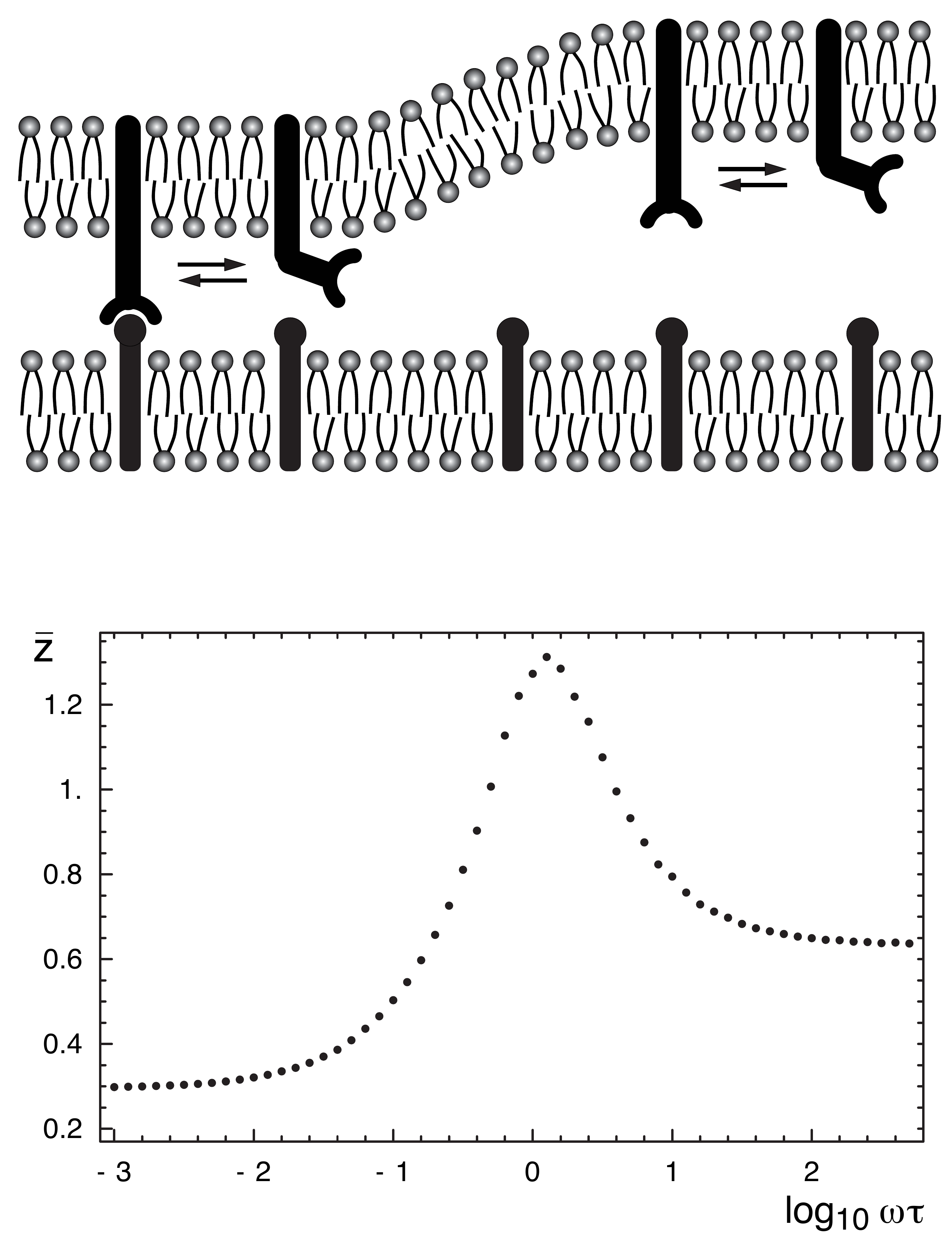}}
\caption{(Top) A membrane with switchable receptors adhering to a second membrane with complementary ligands. The receptors are switched between a stretched, active conformation and a bent, inactive conformation. In the stretched conformation, the adhesion molecules can bind to their ligands in the apposing membrane. --
(Bottom) Monte Carlo data for the average rescaled membrane separation $\bar{z} = \bar{l}/a \sqrt{\kappa/k_B T}$ as a function of the switching rate $\omega=\omega_{+}=\omega_{-}$ of the receptors. Here, $\omega_{+}$ and $\omega_{-}$ are the on- and off-switching rates of the receptors, and $\tau$ is the characteristic relaxation time of a membrane segment with a linear size equal to the mean distance of the receptors.  
The active switching leads to a stochastic resonance with increased membrane separations at intermediate switching rates. The details of the Monte Carlo simulations are described in Ref.~\cite{Rozycki06a}. In this example, the binding energy of the receptors and ligands is $U=2.8 k_B T$.
}
\label{figure_active_switching}
\end{center}
\end{figure}

In the absence of active processes, the adhesiveness of two membranes with complementary receptor and ligand molecules depends on the concentration and binding energies of the molecules, and can be captured by effective adhesion potentials, see section \ref{section_effective_adhesion_potential}. The adhesiveness of membranes with actively switched receptors, in contrast, depends also on the switching rates of the receptors. In the example illustrated in fig.~\ref{figure_active_switching},  the anchored receptors can be switched between two states: an extended `on'-state in which the receptors can bind to ligands anchored in the apposing membrane, and an `off'-state  in which the receptors can't bind. In this example, the switching process from the on- to the off-state requires energy input, e.g.~in the form of ATP. As a consequence, the rate $\omega_{-}$ for this process does not depend on wether the receptor is bound or not, in contrast to equilibrium situations without energy input. In an equilibrium situation, the rate for the transition from the on- to the off-state depends on the binding state and binding energy of a receptor. 

The active switching of the receptors enhances the shape fluctuations of the membranes \cite{Rozycki06a,Rozycki06b,Rozycki07}. Since the steric repulsion of the membranes increases with the shape fluctuations, this enhancement of shape fluctuations leads to larger membrane separations.  In Fig.~\ref{figure_active_switching}(b), the average membrane separation is shown as a function of the switching rates for equal on- and off-rates  $\omega_{+}=\omega_{-}$. In this example, the fractions of receptors in the on- and off-state are constant and equal to $0.5$. The active switching leads to a stochastic resonance of the membrane shape fluctuations, with a maximum of the membrane separation at intermediate switching rates.  At the resonance point, the switching rates are close to the fluctuation relaxation rate $1/\tau$ of a membrane segment with a linear size equal to the average separation of the receptors \cite{Rozycki06a}. 

\section{Discussion and conclusions}

We have reviewed theoretical models for the adhesion of biomimetic membranes and cells {\em via} anchored but mobile receptor and ligand molecules. In these models, the membranes are described as elastic surfaces, and the receptors and ligands as single molecules.
We have argued in the introduction that the elasticity of the membranes is dominated by their bending energy on the relevant lateral length scales up to average separation of the receptor-ligand complexes, which is between 50 and 100 nm for typical concentrations of the complexes in cell adhesion zones \cite{Grakoui99}. The crossover length $\sqrt{\kappa/\sigma}$, above which the tension $\sigma$ dominates over the bending rigidity $\kappa$, is clearly larger for typical membrane tensions $\sigma$, see introduction. However, the average separation of cytoskeletal anchors in cell membranes may be close to the average separation of the receptor-ligand complexes. In the absence of active processes, the coupling of the membrane to the cytoskeleton may lead to a suppression of membrane shape fluctuations on length scales larger than the average separation of the anchors. In the presence of active cytoskeletal processes, the membrane shape fluctuations may even be increased  \cite{Gov03,Auth07}. In the models reviewed here, the cytoskeletal elasticity is neglected since the relevant lateral length scales up to 50 or 100 nm are taken to be smaller than the average separation of the cytoskeletal anchors. However, the active transport of T cell receptors {\em via} a weak coupling to the cytoskeleton has been taken into account in section \ref{section_domain_patterns}, and the active switching of receptors has been considered in section \ref{section_active_switching}.
The characterization of the membrane elasticity by a uniform bending rigidity $\kappa$ is justified on length scales larger than the molecular components of the membranes, i.e.~on length scales larger than 5 or 10 nm. Molecular inhomogeneities within the membranes average out on these length scales, and the presence of anchored or embedded proteins leads to an increased bending rigidity, compared to pure lipid bilayers. 

Important length scales in the direction perpendicular to the membranes are the length of the receptor-ligand complexes, and the binding range of the receptors and ligands. The binding range is the difference between the smallest and largest local membrane separation at which the molecules can bind, and depends on the interaction range of the molecular groups that stick together, the flexibility of the receptor and ligand molecules, and the flexibility of the membrane anchoring. In principle, the binding range may be measured experimentally, or inferred from simulations with atomistic membrane models in a multi-scale modeling approach.  An important quantity is the fraction $P_b$ of the membranes with a local separation within the binding range of the receptors and ligands, see section \ref{section_binding_cooperativity}. The membrane fraction $P_b$ depends on the membrane shape fluctuations on the relevant nanoscales, and thus on the concentrations of the receptors and ligands, which constrain the shape fluctuations as bound complexes. The dependence of $P_b$ on the molecular concentrations leads to cooperative binding. 

As reviewed in section \ref{section_effective_adhesion_potential}, the integration over all possible positions of the receptor and ligand molecules in the partition function of the models leads to effective adhesion potentials for the membranes. These effective adhesion potentials greatly simplify the characterization of the adhesion behavior. In the case of long and short receptor-ligand complexes, for example, the effective adhesion potential allows a general characterization of the critical point for phase separation, see section \ref{subsection_critical_concentrations}. If the adhesion is mediated by adsorbed particles, a similar integration over the degrees of freedom of these particles leads to an effective adhesion energy that is maximal at an optimal particle concentration, see section \ref{section_adsorbed_particles}.


\begin{thebibliography}{118}
\providecommand{\natexlab}[1]{#1}

\bibitem{Alberts02}
Alberts, B., A.~Johnson, J.~Lewis, M.~Raff, K.~Roberts, and P.~Walter. 2002.
\newblock Molecular Biology of the Cell, 4th Ed. Garland, New York.

\bibitem{Alon95}
Alon, R., D.~A. Hammer, and T.~A. Springer. 1995.
\newblock {Lifetime of the P-selectin-carbohydrate bond and its response to
  tensile force in hydrodynamic flow}.
\newblock \emph{Nature}. 374:539--542.

\bibitem{Grakoui99}
Grakoui, A., S.~K. Bromley, C.~Sumen, M.~M. Davis, A.~S. Shaw, P.~M. Allen, and
  M.~L. Dustin. 1999.
\newblock {The immunological synapse: a molecular machine controlling T cell
  activation}.
\newblock \emph{Science}. 285:221--227.

\bibitem{Delanoe04}
Delanoe-Ayari, H., R.~Al~Kurdi, M.~Vallade, D.~Gulino-Debrac, and D.~Riveline.
  2004.
\newblock Membrane and acto-myosin tension promote clustering of adhesion
  proteins.
\newblock \emph{Proc. Natl. Acad. Sci. USA}. 101:2229--2234.

\bibitem{Arnold04}
Arnold, M., E.~A. Cavalcanti-Adam, R.~Glass, J.~Blummel, W.~Eck, M.~Kantlehner,
  H.~Kessler, and J.~P. Spatz. 2004.
\newblock Activation of integrin function by nanopatterned adhesive interfaces.
\newblock \emph{{ChemPhysChem}}. 5:383--388.

\bibitem{Mossman05}
Mossman, K.~D., G.~Campi, J.~T. Groves, and M.~L. Dustin. 2005.
\newblock {Altered TCR signaling from geometrically repatterned immunological
  synapses}.
\newblock \emph{Science}. 310:1191--1193.

\bibitem{Albersdoerfer97}
Albersd\"orfer, A., T.~Feder, and E.~Sackmann. 1997.
\newblock Adhesion-induced domain formation by interplay of long-range
  repulsion and short-range attraction force: a model membrane study.
\newblock \emph{Biophys. J.} 73:245--257.

\bibitem{Maier01}
Maier, C., A.~Behrisch, A.~Kloboucek, D.~Simson, and R.~Merkel. 2001.
\newblock Specific biomembrane adhesion - indirect lateral interactions between
  bound receptor molecules.
\newblock \emph{Eur. Phys. J. E}. 6:273--276.

\bibitem{Smith08}
Smith, A.-S., K.~Sengupta, S.~Goennenwein, U.~Seifert, and E.~Sackmann. 2008.
\newblock Force-induced growth of adhesion domains is controlled by receptor
  mobility.
\newblock \emph{Proc. Natl. Acad. Sci. USA}. 105:6906--6911.

\bibitem{Monks98}
Monks, C.~R., B.~A. Freiberg, H.~Kupfer, N.~Sciaky, and A.~Kupfer. 1998.
\newblock {Three-dimensional segregation of supramolecular activation clusters
  in T cells}.
\newblock \emph{Nature}. 395:82--86.

\bibitem{Davis04}
Davis, D.~M., and M.~L. Dustin. 2004.
\newblock What is the importance of the immunological synapse?
\newblock \emph{Trends. Immunol.} 25:323--327.

\bibitem{Lipowsky96}
Lipowsky, R. 1996.
\newblock Adhesion of membranes via anchored stickers.
\newblock \emph{Phys. Rev. Lett.} 77:1652--1655.

\bibitem{Weikl00}
Weikl, T.~R., R.~R. Netz, and R.~Lipowsky. 2000.
\newblock Unbinding transitions and phase separation of multicomponent
  membranes.
\newblock \emph{Phys. Rev. E.} 62:R45--R48.

\bibitem{Weikl04}
Weikl, T.~R., and R.~Lipowsky. 2004.
\newblock Pattern formation during T-cell adhesion.
\newblock \emph{Biophys. J.} 87:3665--3678.

\bibitem{Asfaw06}
Asfaw, M., B.~R\'{o}\.{z}ycki, R.~Lipowsky, and T.~R. Weikl. 2006.
\newblock Membrane adhesion via competing receptor/ligand bonds.
\newblock \emph{Europhys. Lett.} 76:703--709.

\bibitem{Weikl06}
Weikl, T.~R., and R.~Lipowsky. 2006.
\newblock {Membrane adhesion and domain formation. {\it In} Advances in Planar
  Lipid Bilayers and Liposomes. A. Leitmannova Liu, editor}. Academic Press.

\bibitem{Dustin00}
Dustin, M.~L., and J.~A. Cooper. 2000.
\newblock {The immunological synapse and the actin cytoskeleton: molecular
  hardware for T cell signaling}.
\newblock \emph{Nat. Immunol.} 1:23--29.

\bibitem{Jeppesen01}
Jeppesen, C., J.~Y. Wong, T.~L. Kuhl, J.~N. Israelachvili, N.~Mullah,
  S.~Zalipsky, and C.~M. Marques. 2001.
\newblock Impact of polymer tether length on multiple ligand-receptor bond
  formation.
\newblock \emph{Science}. 293:465--468.

\bibitem{Morreira03}
Moreira, A.~G., C.~Jeppesen, F.~Tanaka, and C.~M. Marques. 2003.
\newblock {Irreversible vs. reversible bridging: When is kinetics relevant for
  adhesion?}
\newblock \emph{Europhys. Lett.} 62:876--882.

\bibitem{Moore06}
Moore, N.~W., and T.~L. Kuhl. 2006.
\newblock The role of flexible tethers in multiple ligand-receptor bond
  formation between curved surfaces.
\newblock \emph{Biophys. J.} 91:1675--1687.

\bibitem{Martin06}
Martin, J.~I., C.~Z. Zhang, and Z.~G. Wang. 2006.
\newblock Polymer-tethered ligand-receptor interactions between surfaces.
\newblock \emph{J. Polym. Sci. B}. 44.

\bibitem{Helfrich73}
Helfrich, W. 1973.
\newblock Elastic properties of lipid bilayers: theory and possible
  experiments.
\newblock \emph{Z. Naturforsch. C}. 28:693--703.

\bibitem{Gov03}
Gov, N., A.~G. Zilman, and S.~Safran. 2003.
\newblock Cytoskeleton confinement and tension of red blood cell membranes.
\newblock \emph{Phys. Rev. Lett.} 90:228101.

\bibitem{Fournier04}
Fournier, J.-B., D.~Lacoste, and E.~Raphael. 2004.
\newblock Fluctuation spectrum of fluid membranes coupled to an elastic
  meshwork: jump of the effective surface tension at the mesh size.
\newblock \emph{Phys. Rev. Lett.} 92:018102.

\bibitem{Lin04}
Lin, L. C.-L., and F.~L.~H. Brown. 2004.
\newblock Dynamics of pinned membranes with application to protein diffusion on
  the surface of red blood cells.
\newblock \emph{Biophys. J.} 86:764--780.

\bibitem{Auth07}
Auth, T., S.~A. Safran, and N.~S. Gov. 2007.
\newblock Fluctuations of coupled fluid and solid membranes with application to
  red blood cells.
\newblock \emph{Phys. Rev. E}. 76:051910.

\bibitem{Lipowsky95}
Lipowsky, R. 1995.
\newblock {Generic interactions of flexible membranes. {\it In} Handbook of
  Biological Physics, Vol. 1. R. Lipowsky and E. Sackmann, editors}.
  Elsevier/North Holland.

\bibitem{Krobath07}
Krobath, H., G.~J. Sch\"utz, R.~Lipowsky, and T.~R. Weikl. 2007.
\newblock {Lateral diffusion of receptor-ligand bonds in membrane adhesion
  zones: Effect of thermal membrane roughness}.
\newblock \emph{Europhys. Lett.} 78:38003.

\bibitem{Selhuber08}
Selhuber-Unkel, C., M.~Lopez-Garcia, H.~Kessler, and J.~Spatz. 2008 Dec.
\newblock Cooperativity in adhesion cluster formation during initial cell
  adhesion.
\newblock \emph{Biophys. J.} 95:5424--5431.

\bibitem{Geiger01}
Geiger, B., and A.~Bershadsky. 2001 Oct.
\newblock Assembly and mechanosensory function of focal contacts.
\newblock \emph{Curr. Opin. Cell Biol.} 13:584--592.

\bibitem{Discher05}
Discher, D.~E., P.~Janmey, and Y.-L. Wang. 2005.
\newblock Tissue cells feel and respond to the stiffness of their substrate.
\newblock \emph{Science}. 310:1139--1143.

\bibitem{Bershadsky06}
Bershadsky, A., M.~Kozlov, and B.~Geiger. 2006.
\newblock Adhesion-mediated mechanosensitivity: a time to experiment, and a
  time to theorize.
\newblock \emph{Curr. Opin. Cell Biol.} 18:472--481.

\bibitem{Girard07}
Girard, P.~P., E.~A. Cavalcanti-Adam, R.~Kemkemer, and J.~P. Spatz. 2007.
\newblock Cellular chemomechanics at interfaces: sensing, integration and
  response.
\newblock \emph{Soft Matter}. 3:307--326.

\bibitem{Schwarz07}
Schwarz, U.~S. 2007.
\newblock Soft matters in cell adhesion: rigidity sensing on soft elastic
  substrates.
\newblock \emph{Soft Matter}. 3.

\bibitem{De08}
De, R., A.~Zemel, and S.~A. Safran. 2008.
\newblock {Do cells sense stress or strain? Measurement of cellular orientation
  can provide a clue}.
\newblock \emph{Biophys. J.} 94:L29--31.

\bibitem{DeMond08}
DeMond, A.~L., K.~D. Mossman, T.~Starr, M.~L. Dustin, and J.~T. Groves. 2008.
\newblock T cell receptor microcluster transport through molecular mazes
  reveals mechanism of translocation.
\newblock \emph{Biophys. J.} 94:3286--3292.

\bibitem{Schuetz00}
Sch\"utz, G.~J., G.~Kada, V.~P. Pastushenko, and H.~Schindler. 2000.
\newblock Properties of lipid microdomains in a muscle cell membrane visualized
  by single molecule microscopy.
\newblock \emph{EMBO J.} 19:892--901.

\bibitem{Sako00}
Sako, Y., K.~Hibino, T.~Miyauchi, Y.~Miyamoto, M.~Ueda, and T.~Yanagida. 2000.
\newblock Single- molecule imaging of signaling molecules in living cells.
\newblock \emph{Single Molecules}. 1:159--163.

\bibitem{Bell78}
Bell, G.~I. 1978.
\newblock Models for the specific adhesion of cells to cells.
\newblock \emph{Science}. 200:618--627.

\bibitem{Bell84}
Bell, G.~I., M.~Dembo, and P.~Bongrand. 1984.
\newblock {Cell adhesion. Competition between nonspecific repulsion and
  specific bonding}.
\newblock \emph{Biophys. J.} 45:1051--1064.

\bibitem{Komura00}
Komura, S., and D.~Andelman. 2000.
\newblock Adhesion-induced lateral phase separation in membranes.
\newblock \emph{Eur. Phys. J. E}. 3:259--271.

\bibitem{Bruinsma00}
Bruinsma, R., A.~Behrisch, and E.~Sackmann. 2000.
\newblock Adhesive switching of membranes: experiment and theory.
\newblock \emph{Phys. Rev. E}. 61:4253--4267.

\bibitem{Chen03}
Chen, H.-Y. 2003.
\newblock Adhesion-induced phase separation of multiple species of membrane
  junctions.
\newblock \emph{Phys. Rev. E}. 67:031919.

\bibitem{Coombs04}
Coombs, D., M.~Dembo, C.~Wofsy, and B.~Goldstein. 2004.
\newblock Equilibrium thermodynamics of cell-cell adhesion mediated by multiple
  ligand-receptor pairs.
\newblock \emph{Biophys. J.} 86:1408--1423.

\bibitem{Shenoy05}
Shenoy, V.~B., and L.~B. Freund. 2005.
\newblock Growth and shape stability of a biological membrane adhesion complex
  in the diffusion-mediated regime.
\newblock \emph{Proc. Natl. Acad. Sci. USA}. 102:3213--3218.

\bibitem{Wu06}
Wu, J.-Y., and H.-Y. Chen. 2006.
\newblock Membrane-adhesion-induced phase separation of two species of
  junctions.
\newblock \emph{Phys. Rev. E}. 73:011914.

\bibitem{Qi01}
Qi, S.~Y., J.~T. Groves, and A.~K. Chakraborty. 2001.
\newblock Synaptic pattern formation during cellular recognition.
\newblock \emph{Proc. Natl. Acad. Sci. USA}. 98:6548--6553.

\bibitem{Raychaudhuri03}
Raychaudhuri, S., A.~K. Chakraborty, and M.~Kardar. 2003.
\newblock Effective membrane model of the immunological synapse.
\newblock \emph{Phys. Rev. Lett.} 91:208101.

\bibitem{Burroughs02}
Burroughs, N.~J., and C.~Wulfing. 2002.
\newblock Differential segregation in a cell-cell contact interface: the
  dynamics of the immunological synapse.
\newblock \emph{Biophys. J.} 83:1784--1796.

\bibitem{Weikl01}
Weikl, T.~R., and R.~Lipowsky. 2001.
\newblock Adhesion-induced phase behavior of multicomponent membranes.
\newblock \emph{Phys. Rev. E.} 64:011903.

\bibitem{Weikl02b}
Weikl, T.~R., D.~Andelman, S.~Komura, and R.~Lipowsky. 2002.
\newblock Adhesion of membranes with competing specific and generic
  interactions.
\newblock \emph{Eur. Phys. J. E}. 8:59--66.

\bibitem{Smith05}
Smith, A.-S., and U.~Seifert. 2005.
\newblock Effective adhesion strength of specifically bound vesicles.
\newblock \emph{Phys. Rev. E}. 71:061902.

\bibitem{Tsourkas07}
Tsourkas, P.~K., N.~Baumgarth, S.~I. Simon, and S.~Raychaudhuri. 2007.
\newblock {Mechanisms of B-cell synapse formation predicted by Monte Carlo
  simulation}.
\newblock \emph{Biophys. J.} 92:4196--4208.

\bibitem{Tsourkas08}
Tsourkas, P.~K., M.~L. Longo, and S.~Raychaudhuri. 2008.
\newblock {Monte Carlo study of single molecule diffusion can elucidate the
  mechanism of B cell synapse formation}.
\newblock \emph{Biophys. J.} 95:1118--1125.

\bibitem{Reister08}
Reister-Gottfried, E., K.~Sengupta, B.~Lorz, E.~Sackmann, U.~Seifert, and A.~S.
  Smith. 2008.
\newblock Dynamics of specific vesicle-substrate adhesion: From local events to
  global dynamics.
\newblock \emph{Phys. Rev. Lett.} 101:208103.

\bibitem{Weikl02a}
Weikl, T.~R., J.~T. Groves, and R.~Lipowsky. 2002.
\newblock Pattern formation during adhesion of multicomponent membranes.
\newblock \emph{Europhys. Lett.} 59:916--922.

\bibitem{Goetz99}
Goetz, R., G.~Gompper, and R.~Lipowsky. 1999.
\newblock Mobilitiy and elasticity of self-assembled membranes.
\newblock \emph{Phys. Rev. Lett.} 82:211--224.

\bibitem{KrobathPreprint}
Krobath, H., B.~R\'{o}\.{z}ycki, R.~Lipowsky, and T.~R. Weikl. 2009.
\newblock Binding cooperativity of membrane adhesion receptors.
\newblock \emph{submitted}. .

\bibitem{Israelachvili92}
Israelachvili, J.~N. 1992.
\newblock Intermolecular and surface forces, 2nd ed. {Academic Press}.

\bibitem{Bayas07}
Bayas, M.~V., A.~Kearney, A.~Avramovic, P.~A. van~der Merwe, and D.~E.
  Leckband. 2007.
\newblock {Impact of salt bridges on the equilibrium binding and adhesion of
  human CD2 and CD58}.
\newblock \emph{J. Biol. Chem.} 282:5589--5596.

\bibitem{Schuck97}
Schuck, P. 1997.
\newblock Use of surface plasmon resonance to probe the equilibrium and dynamic
  aspects of interactions between biological macromolecules.
\newblock \emph{Annu. Rev. Biophys. Biomol. Struct.} 26:541--566.

\bibitem{Rich00}
Rich, R.~L., and D.~G. Myszka. 2000.
\newblock Advances in surface plasmon resonance biosensor analysis.
\newblock \emph{Curr. Opin. Biotechnol.} 11:54--61.

\bibitem{McDonnell01}
McDonnell, J.~M. 2001.
\newblock Surface plasmon resonance: towards an understanding of the mechanisms
  of biological molecular recognition.
\newblock \emph{Curr. Opin. Chem. Biol.} 5:572--577.

\bibitem{Orsello01}
Orsello, C.~E., D.~A. Lauffenburger, and D.~A. Hammer. 2001.
\newblock Molecular properties in cell adhesion: a physical and engineering
  perspective.
\newblock \emph{Trends Biotechnol.} 19:310--316.

\bibitem{Dustin01}
Dustin, M.~L., S.~K. Bromley, M.~M. Davis, and C.~Zhu. 2001.
\newblock Identification of self through two-dimensional chemistry and
  synapses.
\newblock \emph{Annu. Rev. Cell Dev. Biol.} 17:133--157.

\bibitem{Williams01}
Williams, T.~E., S.~Nagarajan, P.~Selvaraj, and C.~Zhu. 2001.
\newblock Quantifying the impact of membrane microtopology on effective
  two-dimensional affinity.
\newblock \emph{J. Biol. Chem.} 276:13283--13288.

\bibitem{Dustin96}
Dustin, M.~L., L.~M. Ferguson, P.~Y. Chan, T.~A. Springer, and D.~E. Golan.
  1996.
\newblock {Visualization of CD2 interaction with LFA-3 and determination of the
  two-dimensional dissociation constant for adhesion receptors in a contact
  area}.
\newblock \emph{J. Cell. Biol.} 132:465--474.

\bibitem{Dustin97}
Dustin, M.~L., D.~E. Golan, D.~M. Zhu, J.~M. Miller, W.~Meier, E.~A. Davies,
  and P.~A. van~der Merwe. 1997.
\newblock {Low affinity interaction of human or rat T cell adhesion molecule
  CD2 with its ligand aligns adhering membranes to achieve high physiological
  affinity}.
\newblock \emph{J. Biol. Chem.} 272:30889--30898.

\bibitem{Zhu07}
Zhu, D.-M., M.~L. Dustin, C.~W. Cairo, and D.~E. Golan. 2007.
\newblock Analysis of two-dimensional dissociation constant of laterally mobile
  cell adhesion molecules.
\newblock \emph{Biophys. J.} 92:1022--1034.

\bibitem{Tolentino08}
Tolentino, T.~P., J.~Wu, V.~I. Zarnitsyna, Y.~Fang, M.~L. Dustin, and C.~Zhu.
  2008.
\newblock {Measuring diffusion and binding kinetics by contact area FRAP}.
\newblock \emph{Biophys. J.} 95:920--930.

\bibitem{Chesla98}
Chesla, S.~E., P.~Selvaraj, and C.~Zhu. 1998.
\newblock {Measuring two-dimensional receptor-ligand binding kinetics by
  micropipette}.
\newblock \emph{Biophys. J.} 75:1553--1572.

\bibitem{Huang04}
Huang, J., J.~Chen, S.~E. Chesla, T.~Yago, P.~Mehta, R.~P. McEver, C.~Zhu, and
  M.~Long. 2004.
\newblock Quantifying the effects of molecular orientation and length on
  two-dimensional receptor-ligand binding kinetics.
\newblock \emph{J. Biol. Chem.} 279:44915--44923.

\bibitem{Kloboucek99}
Kloboucek, A., A.~Behrisch, J.~Faix, and E.~Sackmann. 1999.
\newblock {Adhesion-induced receptor segregation and adhesion plaque formation:
  A model membrane study}.
\newblock \emph{Biophys. J.} 77:2311--2328.

\bibitem{Smith06}
Smith, A.-S., B.~G. Lorz, U.~Seifert, and E.~Sackmann. 2006.
\newblock Antagonist-induced deadhesion of specifically adhered vesicles.
\newblock \emph{Biophys. J.} 90:1064--1080.

\bibitem{Lorz07}
Lorz, B.~G., A.-S. Smith, C.~Gege, and E.~Sackmann. 2007.
\newblock {Adhesion of giant vesicles mediated by weak binding of
  Sialyl-Lewis(x) to E-selectin in the presence of repelling poly(ethylene
  glycol) molecules}.
\newblock \emph{Langmuir}. 23:12293--12300.

\bibitem{Purrucker07}
Purrucker, O., S.~Goennenwein, A.~Foertig, R.~Jordan, M.~Rusp, M.~Baermann,
  L.~Moroder, E.~Sackmann, and M.~Tanaka. 2007.
\newblock Polymer-tethered membranes as quantitative models for the study of
  integrin-mediated cell adhesion.
\newblock \emph{Soft Matter}. 3:333--336.

\bibitem{Lipowsky05}
Lipowsky, R., M.~Brinkmann, R.~Dimova, T.~Franke, J.~Kierfeld, and X.~Zhang.
  2005.
\newblock Droplets, bubbles, and vesicles at chemically structured surfaces.
\newblock \emph{J. Phys.: Condens. Matter}. 17:S537---S558.

\bibitem{Seifert90}
Seifert, U., and R.~Lipowsky. 1990.
\newblock Adhesion of vesicles.
\newblock \emph{Phys. Rev. A}. 42:4768--4771.

\bibitem{Tordeux02}
Tordeux, C., J.-B. Fournier, and P.~Galatola. 2002.
\newblock Analytical characterization of adhering vesicles.
\newblock \emph{Phys. Rev. E}. 65:041912.

\bibitem{Gruhn05}
Gruhn, T., and R.~Lipowsky. 2005.
\newblock Temperature dependence of vesicle adhesion.
\newblock \emph{Phys. Rev. E}. 71:011903.

\bibitem{Lee03}
Lee, S.-J.~E., Y.~Hori, and A.~K. Chakraborty. 2003.
\newblock {Low T cell receptor expression and thermal fluctuations contribute
  to formation of dynamic multifocal synapses in thymocytes}.
\newblock \emph{Proc. Natl. Acad. Sci. USA}. 100:4383--4388.

\bibitem{Figge06}
Figge, M.~T., and M.~Meyer-Hermann. 2006.
\newblock Geometrically repatterned immunological synapses uncover formation
  mechanisms.
\newblock \emph{PLoS Comput. Biol.} 2:e171.

\bibitem{Dustin98}
Dustin, M.~L., M.~W. Olszowy, A.~D. Holdorf, J.~Li, S.~Bromley, N.~Desai,
  P.~Widder, F.~Rosenberger, P.~A. van~der Merwe, P.~M. Allen, and A.~S. Shaw.
  1998.
\newblock A novel adaptor protein orchestrates receptor patterning and
  cytoskeletal polarity in t-cell contacts.
\newblock \emph{Cell}. 94:667--677.

\bibitem{footnote}Time scales for adhesion are obtained by comparing the diffusion constants of the receptors and ligands with experimental values, see ref.~\cite{Weikl04} for details. 

\bibitem{Janeway07}
Murphy, K.~M., P.~Travers, and M.~Walport. 2007.
\newblock Janeway's Immunobiology, 7th ed. Garland Science.

\bibitem{Choudhuri07}
Choudhuri, K., and P.~A. van~der Merwe. 2007.
\newblock {Molecular mechanisms involved in T cell receptor triggering}.
\newblock \emph{Semin. Immunol.} 19:255--261.

\bibitem{Davis06}
Davis, S.~J., and P.~A. van~der Merwe. 2006 Aug.
\newblock {The kinetic-segregation model: TCR triggering and beyond}.
\newblock \emph{Nat. Immunol.} 7:803--809.

\bibitem{Krogsgaard07}
Krogsgaard, M., J.~Juang, and M.~M. Davis. 2007.
\newblock {A role for "self" in T-cell activation}.
\newblock \emph{Semin. Immunol.} 19:236--244.

\bibitem{Campi05}
Campi, G., R.~Varma, and M.~Dustin. 2005.
\newblock {Actin and agonist MHC-peptide complex-dependent T cell receptor
  microclusters as scaffolds for signaling}.
\newblock \emph{J. Exp. Med.} 202:1031--1036.

\bibitem{Yokosuka05}
Yokosuka, T., K.~Sakata-Sogawa, W.~Kobayashi, M.~Hiroshima, A.~Hashimoto-Tane,
  M.~Tokunaga, M.~Dustin, and T.~Saito. 2005.
\newblock {Newly generated T cell receptor microclusters initiate and sustain T
  cell activation by recruitment of Zap70 and SLP-76}.
\newblock \emph{Nat. Immunol.} 6:1253--1262.

\bibitem{Bunnell02}
Bunnell, S.~C., D.~I. Hong, J.~R. Kardon, T.~Yamazaki, C.~J. McGlade, V.~A.
  Barr, and L.~E. Samelson. 2002.
\newblock {T cell receptor ligation induces the formation of dynamically
  regulated signaling assemblies}.
\newblock \emph{J. Cell. Biol.} 158:1263--1275.

\bibitem{Varma06}
Varma, R., G.~Campi, T.~Yokosuka, T.~Saito, and M.~L. Dustin. 2006.
\newblock T cell receptor-proximal signals are sustained in peripheral
  microclusters and terminated in the central supramolecular activation
  cluster.
\newblock \emph{Immunity}. 25:117--127.

\bibitem{Yokosuka08}
Yokosuka, T., W.~Kobayashi, K.~Sakata-Sogawa, M.~Takamatsu, A.~Hashimoto-Tane,
  M.~L. Dustin, M.~Tokunaga, and T.~Saito. 2008.
\newblock {Spatiotemporal regulation of T cell costimulation by TCR-CD28
  microclusters and protein kinase C theta translocation}.
\newblock \emph{Immunity}. 29:589--601.

\bibitem{Yokosuka09}
Yokosuka, T., and T.~Saito. 2009.
\newblock {Dynamic regulation of T-cell costimulation through TCR-CD28
  microclusters}.
\newblock \emph{Immunol. Rev.} 229:27--40.

\bibitem{Irvine02}
Irvine, D.~J., M.~A. Purbhoo, M.~Krogsgaard, and M.~M. Davis. 2002.
\newblock {Direct observation of ligand recognition by T cells}.
\newblock \emph{Nature}. 419:845--849.

\bibitem{Purbhoo04}
Purbhoo, M.~A., D.~J. Irvine, J.~B. Huppa, and M.~M. Davis. 2004.
\newblock T cell killing does not require the formation of a stable mature
  immunological synapse.
\newblock \emph{Nat. Immunol.} 5:524--530.

\bibitem{Leckband94}
Leckband, D.~E., F.~J. Schmitt, J.~N. Israelachvili, and W.~Knoll. 1994.
\newblock Direct force measurements of specific and nonspecific protein
  interactions.
\newblock \emph{Biochemistry}. 33:4611--4624.

\bibitem{Hu04}
Hu, Y., I.~Doudevski, D.~Wood, M.~Moscarello, C.~Husted, C.~Genain, J.~A.
  Zasadzinski, and J.~Israelachvili. 2004.
\newblock Synergistic interactions of lipids and myelin basic protein.
\newblock \emph{Proc. Natl. Acad. Sci. USA}. 101:13466--13471.

\bibitem{Baksh04}
Baksh, M.~M., M.~Jaros, and J.~T. Groves. 2004.
\newblock Detection of molecular interactions at membrane surfaces through
  colloid phase transitions.
\newblock \emph{Nature}. 427:139--141.

\bibitem{Winter06}
Winter, E.~M., and J.~T. Groves. 2006.
\newblock Surface binding affinity measurements from order transitions of lipid
  membrane-coated colloidal particles.
\newblock \emph{Anal. Chem.} 78:174--180.

\bibitem{Rozycki08b}
R\'{o}\.{z}ycki, B., R.~Lipowsky, and T.~R. Weikl. 2008.
\newblock Effective surface interactions mediated by adhesive particles.
\newblock \emph{Europhys. Lett.} 84:26004.

\bibitem{Takagi02}
Takagi, J., B.~M. Petre, T.~Walz, and T.~A. Springer. 2002.
\newblock Global conformational rearrangements in integrin extracellular
  domains in outside-in and inside-out signaling.
\newblock \emph{Cell}. 110:599--511.

\bibitem{Kim03}
Kim, M., C.~V. Carman, and T.~A. Springer. 2003.
\newblock Bidirectional transmembrane signaling by cytoplasmic domain
  separation in integrins.
\newblock \emph{Science}. 301:1720--1725.

\bibitem{Dustin04}
Dustin, M., T.~Bivona, and M.~Philips. 2004.
\newblock {Membranes as messengers in T cell adhesion signaling}.
\newblock \emph{Nat. Immunol.} 5:363--372.

\bibitem{Moeller98}
M\"oller, G., M.~Harke, H.~Motschmann, and D.~Prescher. 1998.
\newblock Controlling microdroplet formation by light.
\newblock \emph{Langmuir}. 14:4955--4957.
\newblock Langmuir.

\bibitem{Ichimura00}
Ichimura, K., S.~Oh, and M.~Nakagawa. 2000.
\newblock Light-driven motion of liquids on a photoresponsive surface.
\newblock \emph{Science}. 288:1624--1626.

\bibitem{Prost96}
Prost, J., and R.~Bruinsma. 1996.
\newblock Shape fluctuations of active membranes.
\newblock \emph{Europhys. Lett.} 33:321--326.

\bibitem{Manneville99}
Manneville, J., P.~Bassereau, D.~Levy, and J.~Prost. 1999.
\newblock Activity of transmembrane proteins induces magnification of shape
  fluctuations of lipid membranes.
\newblock \emph{Phys. Rev. Lett.} 82:4356--4359.

\bibitem{Ramaswamy00}
Ramaswamy, S., J.~Toner, and J.~Prost. 2000.
\newblock Nonequilibrium fluctuations, traveling waves, and instabilities in
  active membranes.
\newblock \emph{Phys. Rev. Lett.} 84:3494--3497.

\bibitem{Gov04}
Gov, N. 2004.
\newblock Membrane undulations driven by force fluctuations of active proteins.
\newblock \emph{Phys. Rev. Lett.} 93:268104.

\bibitem{Lin06}
Lin, L. C.-L., N.~Gov, and F.~L.~H. Brown. 2006.
\newblock Nonequilibrium membrane fluctuations driven by active proteins.
\newblock \emph{J. Chem. Phys.} 124:74903.

\bibitem{ElAlaouiFaris09}
El~Alaoui~Faris, M.~D., D.~Lacoste, J.~J.~Pecreaux, J.-F. Joanny, J.~Prost, and
  P.~Bassereau. 2009.
\newblock Membrane tension lowering induced by protein activity.
\newblock \emph{Phys. Rev. Lett.} 102:038102.

\bibitem{Gov05}
Gov, N., and S.~Safran. 2005.
\newblock {Red blood cell membrane fluctuations and shape controlled by
  ATP-induced cytoskeletal defects}.
\newblock \emph{Biophys. J.} 88:1859--1874.

\bibitem{Zhang08b}
Zhang, R., and F.~L.~H. Brown. 2008.
\newblock Cytoskeleton mediated effective elastic properties of model red blood
  cell membranes.
\newblock \emph{J. Chem. Phys.} 129:065101.

\bibitem{Sabra98}
Sabra, M., and O.~Mouritsen. 1998.
\newblock Steady-state compartmentalization of lipid membranes by active
  proteins.
\newblock \emph{Biophys. J.} 74:745--752.

\bibitem{Chen04}
Chen, H. 2004.
\newblock Internal states of active inclusions and the dynamics of an active
  membrane.
\newblock \emph{Phys. Rev. Lett.} 92:168101.

\bibitem{Rozycki06a}
R\'{o}\.{z}ycki, B., R.~Lipowsky, and T.~R. Weikl. 2006{\natexlab{a}}.
\newblock Adhesion of membranes with active stickers.
\newblock \emph{Phys. Rev. Lett.} 96:048101.

\bibitem{Rozycki06b}
R\'{o}\.{z}ycki, B., T.~R. Weikl, and R.~Lipowsky. 2006{\natexlab{b}}.
\newblock Adhesion of membranes via switchable molecules.
\newblock \emph{Phys. Rev. E}. 73:061908.

\bibitem{Rozycki07}
R\'{o}\.{z}ycki, B., T.~R. Weikl, and R.~Lipowsky. 2007.
\newblock Stochastic resonance for adhesion of membranes with active stickers.
\newblock \emph{Eur. Phys. J. E}. 22:97--106.

\end{thebibliography}
\end{document}